\documentclass[aps,prb, notitlepage, preprintnumbers, nofootinbib, twocolumn, 
nolongbibliography,amsmath,amssymb,superscriptaddress,floatfix,scrartcl, longbibliography]{revtex4-2}

\usepackage{amssymb}
\usepackage{color}
\usepackage{graphicx}
\usepackage{mathrsfs}
\usepackage[unicode=true,pdfusetitle,bookmarks=false,colorlinks=true,citecolor=blue,urlcolor=blue,linkcolor=red]{hyperref}
\usepackage{bm}
\usepackage{braket}
\usepackage{tikz}
\usepackage{subfigure, epsfig}
\usetikzlibrary{decorations.markings, calc, arrows.meta, bending}

\usepackage{cleveref}
\crefname{appendix}{App.}{Apps.}
\crefname{equation}{Eq.}{Eqs.}
\crefname{figure}{Fig.}{Figs.}
\crefname{table}{Tab.}{Tabs.}
\crefname{section}{Sec.}{Secs.}

\begin{document}

\title{Multi wavefunction overlap and multi entropy for topological ground states in (2+1) dimensions}
\preprint{RIKEN-iTHEMS-Report-24}
\preprint{KYUSHU-HET-297}

\author{Bowei Liu}
\email{boweil@princeton.edu}
\affiliation{Department of Physics, Princeton University, Princeton, New Jersey 08544, USA}

\author{Junjia Zhang}
\affiliation{Department of Physics, Princeton University, Princeton, New Jersey 08544, USA}

\author{Shuhei Ohyama}
\email{shuhei.ohyama@riken.jp}
\affiliation{
RIKEN Center for Emergent Matter Science, Wako, Saitama, 351-0198, Japan}
\affiliation{\it University of Vienna, Faculty of Physics, Boltzmanngasse 5, A-1090 Vienna, Austria}

\author{Yuya Kusuki}
\affiliation{\it Institute for Advanced Study, Kyushu University, Fukuoka 819-0395, Japan}
\affiliation{\it Department of Physics, Kyushu University, 
Fukuoka 819-0395, Japan}
\affiliation{RIKEN Interdisciplinary Theoretical and Mathematical Sciences (iTHEMS), \\ Wako, Saitama 351-0198, Japan}

\author{Shinsei Ryu}
\affiliation{Department of Physics, Princeton University, Princeton, New Jersey 08544, USA}

\date{\today}

\begin{abstract}
Multi-wavefunction overlaps -- generalizations of the quantum mechanical inner product for more than two quantum many-body states -- are valuable tools for studying many-body physics. In this paper, we investigate the multi-wavefunction overlap of (2+1)-dimensional gapped ground states, focusing particularly on symmetry-protected topological (SPT) states. We demonstrate how these overlaps can be calculated using the bulk-boundary correspondence and (1+1)-dimensional edge theories, specifically conformal field theory. When applied to SPT phases, we show that the topological invariants, which can be thought of as discrete higher Berry phases, can be extracted from the multi-wavefunction overlap of four ground states with appropriate symmetry actions. Additionally, we find that the multi-wavefunction overlap can be expressed in terms of the realignment of reduced density matrices. Furthermore, we illustrate that the same technique can be used to evaluate the multi-entropy -- a quantum information theoretical quantity associated with multi-partition of many-body quantum states
--for (2+1)-dimensional gapped ground states.
Combined with numerics, 
we show that the difference between
multi-entropy for tripartition
and second R\'enyi entropies is bounded from below
by $(c_{{\it tot}}/4)\ln 2$ where 
$c_{{\it tot}}$ is the central charge of 
ungappable degrees of freedom.
To calculate multi-entropy numerically for 
free fermion systems (such as Chern insulators),
we develop the correlator method for multi-entropy.
\end{abstract}

\pacs{??}

\maketitle


\begin{figure*}[ht]
\centering
\includegraphics[scale=1.7]{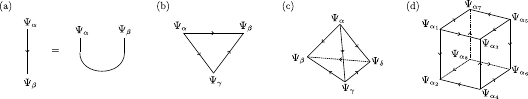}
\caption{
\label{fig1}
(a) The regular quantum mechanical inner product between two states
$\Psi_{\alpha}=\sum_i\psi_{i\alpha}|i\rangle$ and
$\Psi_{\beta}=\sum_i\psi_{i\beta}|i\rangle$,
$\langle \Psi_{\beta} | \Psi_{\alpha}\rangle
= \sum_i \psi^*_{\beta i}\psi_{\alpha i}$.
The inner product can also be written in terms of a maximally entangled state
that is represented as a cup on the right-hand side.
entangled state. 
(b) The triple inner product for three states. Each state is defined in a
bipartite Hilbert space, e.g., $\Psi_{\alpha}=\sum_{i,j}\psi_{ij, \alpha}|i\rangle|j\rangle$.
(c) The quadruple inner product. 
Each state is defined in a
tripartite Hilbert space, e.g., $\Psi_{\alpha}=\sum_{i,j,k}\psi_{ijk, \alpha}|i\rangle|j\rangle|k\rangle$.
(d) The multi-wavefunction overlap relevant to multi-entropy.
}
\end{figure*}


\section{Introduction}

%
%
%
%
%
%


Understanding the pattern of quantum entanglement in many-body quantum states is
key to discovering and diagnosing novel (topological) phases of matter and topological effects.
Developing such understanding is also essential for devising
(numerical) tools to attack quantum many-body problems.
For example, topologically ordered phases of matter are characterized not by local order
parameters but by their patterns of long-range entanglement.
It has been found that the scaling of the entanglement entropy for
bipartition captures a universal property of topological ground states
(the so-called topological entanglement entropy
\cite{2006PhRvL..96k0404K,2006PhRvL..96k0405L}).
Today, topological entanglement entropy is accepted as a smoking gun for
topological ground states, and has been ``measured'' in quantum devices
\cite{satzinger_realizing_2021}.

In recent years, the study of multipartite quantum entanglement and correlations in many-body systems, in contrast to bipartite quantum entanglement, has gained significant attention.
In fact, while topological entanglement entropy, defined in terms of bipartite entanglement, provides some insights into topological liquids, it falls short of fully characterizing them; Different topological states can share the same topological entanglement entropy but remain distinct. Recent studies have begun to explore multipartite entanglement of topological ground states using various entanglement measures such as reflected entropy, entanglement negativity, and others 
\cite{Liu_2022, Siva_2022, liu2023multipartite, Sohal_2023,
Lee_2013,
Castelnovo_2013,2016PhRvB..93x5140W, Wen_2016, 2021PhRvB.104k5155L,
liu2022entanglement,
Lu_2020}.

\begin{figure*}[ht]
  \centering
  \includegraphics[scale=1.7]{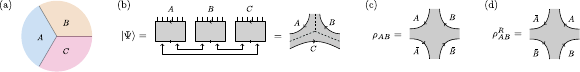}
  \caption{
    \label{tripartition setup}
    (a) The tripartition of a (2+1)d gapped ground state into three regions $A, B, C$.
    (b) The representation of the tripartite ground state using the bulk-boundary
    correspondence and the vertex state in edge conformal field theory.
    (c) The path integral representation of the reduced density matrix $\rho_{AB}$.
    (d) The path integral representation of the realignment $\rho^R_{AB}$.
  }
\end{figure*}

For instance, 
reflected entropy for three spatial regions depicted in Fig.\ \ref{tripartition setup}(a) has been investigated
\cite{Liu_2022, Siva_2022, liu2023multipartite, Sohal_2023}.
Reflected entropy is expected to capture quantum correlations beyond simple Bell- or EPR-type bipartite correlations.
Specifically, it was found that the Markov gap (the difference between the reflected entropy and the mutual information)
for two regions (after tracing out $C$) is independent of the subregion sizes
and bound from below by the universal number,
$(c/3)\ln 2$, where $c$ is the central charge
of stable (ungappable) degrees of freedom of edge conformal field theory.
A non-zero Markov gap can thus detect an obstruction to completely gapping the edge theory.
Reflected entropy has also been studied in conformal field theory and holographic duality
\cite{2021JHEP...03..178D, Akers_2020, 2021JHEP...10..047H};
in fact, 
the concept of reflected entropy is first conceived in the holographic context
\cite{2021JHEP...03..178D}.
See \cite{Shi_2020,
Shi_2020b,Kim_2022b, Fan_2022, Zou_2022c, fan2022extracting} for other relevant works.


In this work, we study the multi-wavefunction overlap and multi-entropy
for gapped ground states in (2+1) dimensions.
The multi-entropy is an entanglement quantity that captures multipartite correlations and entanglement.
It is defined by generalizing the replica trick to symmetric permutations
of multiple sub-Hilbert spaces (subregions)
\cite{Gadde_2022,Penington:2022dhr,Harper:2024ker,Gadde:2023zzj}.
Similar to the holographic entanglement entropy in AdS/CFT,
multi-entropy also has a holographic computation in the bulk geometry,
given by the minimum area of surfaces that tripartition the bulk geometry, divided by 
$4 G_N$ where $G_N$ is the Newton constant.
See Sec.\ \ref{sec:multi} for detailed definitions.

Multi-wavefunction overlaps -- generalizations of the quantum mechanical inner product for more than two quantum many-body wavefunctions 
-- have been introduced and utilized 
in (1+1)d many-body systems 
\cite{liu2022operator,ohyama2023higher,ohyama2024higher_a}.
Notably, it has been demonstrated that multi-wavefunction overlaps can extract the higher Berry phase, which is a many-body generalization of the regular Berry phase in single-particle quantum mechanics 
\cite{Kitaev13,
KS20-1,KS20-2,
Hsin_2020,
Cordova_2020a, Cordova_2020b,
shiozaki21,
Choi_2022,
Wen_2023,
OTS23,
beaudry2023homotopical,
qi2023charting,
shiozaki2023higher,
artymowicz2023quantization,
yao2024modulating,
ohyama2024higher_b,sommer2024higher_b,sommer2024higher_a}.

While these subjects,
multipartite entanglement and the higher Berry phase,
can be studied separately in principle,
they are closely linked conceptually
-- both multi-entropy and multi-wavefunction overlaps
are closely related to multipartite entanglement 
in many-body systems.
%
To briefly explain the connection, 
we note that the regular inner product for two states can also be formulated in terms of a maximally entangled state of two parties
[Fig.\ \ref{fig1}(a)].
It would then be a natural step to generalize the regular inner product 
by introducing quantum states with multipartite entanglement. 
This would result in a generalization of the regular inner product 
that is defined for more than two quantum states.
In this work, we will use the so-called vertex states
in (1+1)d CFT 
\cite{Liu_2022, 
liu2023multipartite}
-- a particular tripartite entangled state
to formulate and calculate multi-wavefunction overlap.
This can be thought of as a multipartite generalization of
the Li-Haldane conjecture
\cite{2006PhRvL..96k0404K, 2008PhRvL.101a0504L, qi2012general,2016PhRvB..93x5140W}.

The objects we consider in this paper are summarized in Fig.\ \ref{fig1};
triple [Fig.\ \ref{fig1}(b)] and quadruple
[Fig.\ \ref{fig1}(c)]
inner products of many-body wavefunctions;
multi-entropy
[Fig.\ \ref{fig1}(d)].
The more precise definitions of these quantities will be discussed later. 

The rest of the paper is organized as follows.
After going through the necessary backgrounds
in Sec.\ \ref{sec:preliminaries}, 
we discuss multi-wavefunction overlaps in 
in Sec.\ \ref{Multi wavefunction overlap}.
To put our calculations in a proper context,
we first review the triple inner product of
(1+1)d short-ranged entangled (invertible) ground states
in Sec.\ \ref{Triple inner product in (1+1)d}
[Fig.\ \ref{fig1}(b)].
In Sec.\ \ref{Quadruple inner product in (2+1)d},
we calculate the quadruple inner product of (2+1)d invertible ground states
[Fig.\ \ref{fig1}(c)].
In our calculations, we use the bulk-boundary correspondence and vertex states reviewed in
Sec.\ \ref{sec:preliminaries}.
In particular, we show that
we can extract the topological invariant of
symmetry-protected topological (SPT) phases
from wavefunction overlaps.
Next,
in Sec.\ \ref{sec:multi},
we calculate multi-entropy for (2+1)d invertible ground states
using tripartition in Fig.\ \ref{tripartition setup}(a) 
[Fig.\ \ref{fig1}(d)].
We close in Sec.\ \ref{Discussion}
with a brief discussion on future issues.

\section{Preliminaries}
\label{sec:preliminaries}

In this section, we go through the necessary ingredients for our calculations
of the multi-wavefunction overlap and multi-entropy.

\subsection{Bulk-boundary correspondence and vertex states}
\label{sec:Bulk-boundary correspondence and vertex states}


The bulk-boundary correspondence is a hallmark of topological phases of matter,
including both long-range (topologically-ordered) and short-range entangled (invertible) states.
It establishes a relationship between the bulk topological properties of gapped ground states and
quantum anomalies of edge excitations that arise at a system's boundary.
In addition to physical boundaries, the bulk-boundary correspondence is also 
applicable to 
entanglement quantities of topological ground states, such as entanglement entropy 
(and its topological part) associated with a given spatial subregion. 
For entanglement entropy, the bipartition of topological ground states leads to
a "virtual" or entangling boundary.
Much like the physical Hamiltonian with a boundary, the reduced density matrix or
the entanglement Hamiltonian of the subregion has a low-lying spectrum supported only near the entangling boundary.
Furthermore, for (2+1)d topological phases, 
this low-lying part of the entanglement Hamiltonian is well captured by the corresponding 
(1+1)d CFT, which also describes the physical edge excitations. 
This line of thinking also allows us to write down the Schmidt decomposition 
of the topological ground state using a conformal boundary state (Ishibashi boundary state).
This boundary state is written in terms of the boundary degree of the region $A$ and its complement $\bar{A}$.
Explicitly, a topological ground state $|\Psi \rangle$ 
is written in terms of the low-lying Schmidt basis as 
$|\Psi\rangle \propto e^{ -\beta H_{{\it edge}}}|B\rangle$
where $H_{{\it edge}}$ is a non-chiral CFT Hamiltonian describing the edge theories 
for the boundaries of $A$ and $\bar{A}$, $|B\rangle$ is a boundary state of the CFT, and
$\beta$ is proportional to the inverse bulk gap.
The reduced density matrix $\rho_A$, obtained by tracing out the degrees of freedom of the edge theory of $\bar{A}$, is 
$\rho_A \propto e^{ - 2\beta H_{{\it edge}}}$, as expected from the bulk-boundary correspondence.

The quantities of our interest, namely, multi-entropy and multi-wavefunction overlaps, 
require tripartition (or multipartition) of topological ground states. 
To this end, we make use of the multipartite generalization of the above setup
introduced in Refs.\ \cite{Liu_2022, liu2023multipartite}.
In short, when considering the tripartition of a topological ground state $|\Psi\rangle$
depicted in Fig.\ \ref{tripartition setup},
we assume that $|\Psi\rangle$ can be well approximated as 
\begin{align}
|\Psi\rangle \propto e^{ - \beta H_{{\it edge}}} |V\rangle
\end{align}
where $H_{{\it edge}} = H_{\partial A} + H_{\partial B} + H_{\partial C}$
represents the Hamiltonian for the edge CFT
at the boundaries of region $A$, $B$ and $C$,
and $\beta$ is again proportional to the inverse bulk gap. 
The state $|V\rangle$, defined in the tensor product Hilbert space 
${\cal H}_{\partial A}\otimes {\cal H}_{\partial B} \otimes {\cal H}_{\partial C}$, 
plays a role analogous to the boundary state $|B\rangle$ for the case of bipartition.
A formal definition can be found in Ref.\ \cite{Liu_2022}.
What is important for us is the path integral representation of 
$e^{ - \beta H_{{\it edge}}} |V\rangle$,
shown in Fig.\ \ref{tripartition setup}(b).
Here, the initial state $|V\rangle$ located at the initial time slice (dotted lines) is
time-evolved by the Euclidean time-evolution operator
$e^{ - \beta H}$ (from now on, we simplify the notation by writing the (1+1)d edge Hamiltonian as $H$).
At the initial time slice, the left half of the edge $\partial A$
is glued or identified with the left half of $\partial B$,
and the right half of $\partial B$
is identified with the left half of $\partial C$, etc., 
as indicated by arrows. 
This specific boundary condition (initial condition) imposed on the path integral 
results from having $|V\rangle$ as the initial condition.
The path integral can also be represented as the right-most diagram in  
Fig.\ \ref{tripartition setup}.
We will frequently use this path integral representation of the state $e^{-\beta H}|V\rangle$.


Several comments are in order:

-- First, vertex states can be defined both under open boundary conditions (OBC) 
and periodic boundary conditions (PBC).
In the following, we will mainly use OBC. 
In terms of the (2+1)d setup, 
the setup of our primary interest 
involves 
the tripartition of topological states put on an infinite plane with {\it one} trijunction where the three regions $A$, $B$, and $C$ meet.
To represent this configuration in the edge CFT language, 
we take a proper limit 
where we send the spatial length of the 1d entangling boundary
($\equiv L$) infinity,
$L/\beta \to \infty$.
We will discuss this limit in more detail in later sections.
On the other hand,
a vertex state with PBC corresponds to the tripartition of a topological state
put on a spatial sphere. 
We should note that such tripartition
has {\it two} trijunctions.
Our setup can also be obtained from this spherical configuration 
by taking a proper limit, sending one of the trijunctions to infinity.

-- Second, in the path integral in Fig.\ \ref{tripartition setup}(b), 
defect lines may appear
along the entangling boundaries [dotted lines in 
 Fig.\ \ref{tripartition setup}(b)].
These defect lines are denoted as ${\cal I}$ and ${\cal I}^{\dag}$
in Ref.\ \cite{liu2023multipartite}.
For topologically-ordered phases, we need 
the projector 
onto a given topological charge 
(single Verma module), 
which results in 
topological entanglement entropy.
For invertible states (e.g., SPT phases)
with no topological entanglement entropy,
there is no projector (interface) in the edge theory.
In other words, since SPT phases are adiabatically deformable to a trivial state,
the edge theory admits a (non-symmetric) Cardy state.
This implies that the edge description of the multipartite state is just given by the regular (possibly higher-genus) partition function.
Therefore, for SPT phase, defect lines are absent.

\begin{figure*}[ht]
\centering
\includegraphics[scale=1.8]{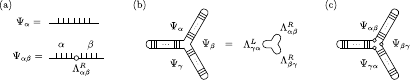}
\caption{
\label{fig:1+1d}
(a) The MPS representations of (1+1)d invertible states $\Psi_{\alpha}$;
$\Psi_{\alpha\beta}$ is a MPS representation in the mixed gauge;
(b) The triple inner product of $\Psi_{\alpha},\Psi_{\beta},\Psi_{\gamma}$;
(c) The triple inner product in the mixed gauge.
}
\end{figure*}

\subsection{Realignment}
\label{sec:realignment}

The quantities of our interest, namely, 
multi-entropy and multi-wavefunction overlap, 
can be expressed using the realignment 
\cite{Rudolph_2003, chen2003matrix, Rudolph_2005}
of reduced density matrices.
Here, we go through the relevant definitions and the path integral representations.

Let us consider a (reduced) density matrix $\rho_{AB}$ on
the bipartite Hilbert space $\mathcal{H}_A\otimes \mathcal{H}_B$.
It can be expanded in the (computational) basis 
$\{|i\rangle_A |j\rangle_B = |ij\rangle\}$
as
$\rho_{A B}=\sum_{i, k=1}^{\mathrm{dim}\, {\cal H}_A} 
\sum_{j, l=1}^{\mathrm{dim}\, {\cal H}_B} \rho_{i j, k l}|i j\rangle\langle k l|$.
Here, $i,k$ label the basis in $\mathcal{H}_A^* \cong \mathcal{H}_A$, and $j,l$ label the indices in $\mathcal{H}_B^* \cong \mathcal{H}_B$. We define the realigned reduced density matrix as $\rho_{A B}^R=\sum_{i, k=1}^{\mathrm{dim}\, {\cal H}_A} 
\sum_{j, l=1}^{\mathrm{dim}\, {\cal H}_B} \rho_{i j, k l}|i k\rangle\langle j l|$. 
Similar to partial transpose, which one can use to develop the corresponding
positive partial transpose (PPT) criterion and a mixed-state entanglement measure
(entanglement negativity) 
\cite{Peres_1996,Horodecki1996,Simon2000,
PhysRevLett.86.3658,PhysRevLett.87.167904,Zyczkowski1,
Zyczkowski2,
PlenioEisert1999,Vidal_2002,Plenio2005},
one can make use of realignment to detect mixed-state entanglement.
Recently, the computable cross norm or realignment (CCNR) negativity has been studied
in many-body systems including topological phases and (conformal) field theories -- see, for example, 
Refs.\ \cite{Yin_2023a,
Yin_2023, Berthiere_2021,
Bruno_2024, milekhin2022computablecrossnormtensor}.

Using the path integral representation of a tripartite gapped topological ground state
$|\Psi\rangle$ in Fig.\ \ref{tripartition setup}(b),
we define realignment as follows in our setup. 
First, we consider the reduced density 
matrix $\rho_{AB}$ by taking partial trace over $C$,
$\rho_{AB}= \mathrm{Tr}_C |\Psi\rangle \langle\Psi|$. 
The path integral representation of $\rho_{AB}$
is depicted in Fig.\ \ref{tripartition setup}(c),
where the arrows indicate the chirality of the edge state on $\partial A$, $\partial B$ and $\partial C$, which is consistent with the orientation of 
the Riemann surface on which the path integral is defined.
$\langle \Psi|$ can be represented similarly.
We then define the realignment of $\rho_{AB}$, $\rho^R_{AB}$,  
by rotating the path integral 
by 90 degrees clockwise, 
as in Fig.\ \ref{tripartition setup}(d).
(Here, the labels $A, B, \bar{A}, \bar{B}$ 
are inherited from the previous picture. They do not represent the Hilbert spaces.
The Hilbert spaces are indicated by their locations.)
Note that in our setup
$\mathrm{dim}\, {\cal H}_A = 
\mathrm{dim}\, {\cal H}_B
=
\mathrm{dim}\, {\cal H}_C
$.


\section{Multi-wavefunction overlap}
\label{Multi wavefunction overlap}

%
%
%
%
%
%
%
In this section, 
we explore the multi-wavefunction overlap of
four many-body ground states in (2+1) dimensions. 
For each wavefunction, we consider tripartition into 
three spatial regions $A,B,C$,
as in Fig.\ \ref{tripartition setup}(a).
We contract the different regions in a given wavefunction with different regions of other wavefunctions (the details will be described below). 
As a specific example, we focus on 
invertible 
(short-range entangled) ground states protected by some symmetry, i.e., 
SPT states.
(It should however be noted that our techniques are not limited to SPT states.)
More precisely, 
we consider states generated from a given invertible state by acting with 
global or partial symmetry transformation, as explained below.

\subsection{Triple inner product in (1+1)d}
\label{Triple inner product in (1+1)d}

To put our calculations in the proper context, 
let us start by reviewing the multi-wavefunction overlap
for (1+1)d gapped ground states (invertible states)  
following Ref.\ \cite{ohyama2023higher}.
Let us consider three invertible ground states $\Psi_{\alpha}$, $\Psi_{\beta}$, and $\Psi_{\gamma}$.
We assume that they are represented by Matrix Product States (MPS)
\cite{Cirac_2021}.
More specifically, 
we consider translationally invariant, normal MPS in the right canonical gauge.
By using the left and right eigenvectors of the transfer matrix,
we take the thermodynamic limit 
and consider the infinite MPS.
We further assume that 
these MPS represent physically the same state,
but are represented by different MPS matrices,
$\{A^s_{\alpha}\}_{s=1,\cdots, \mathsf{D}}$, 
$\{A^s_{\beta}\}_{s=1,\cdots, \mathsf{D}}$,
and
$\{A^s_{\gamma}\}_{s=1,\cdots, \mathsf{D}}
$,
respectively.
(Here, $s=1,\cdots, \mathsf{D}$ represents a physical "spin" index,
while virtual or internal indices are implicit.)
Namely, they are in different MPS gauges. 
By the fundamental theorem, these MPS are related by, e.g., 
$A^s_{\alpha} = g^{\ }_{\alpha\beta} A^{s}_{\beta} g^{\dag}_{\alpha\beta}
$
(and similar relations for any pair of $\alpha,\beta,\gamma$),
where the transition function $g_{\alpha\beta}$
is an element of the projective unitary group.

For these three MPS states, their triple inner product is defined first by considering the bipartition of the space into two regions, $A$ and $B$, say, 
and contract the MPS as in Fig.\ \ref{fig:1+1d}(a).
Here, we note that 
$\Psi_{\alpha}$ is unconjugated,
$\Psi_{\beta}$ is conjugated, and
$\Psi_{\gamma}$ is "partially conjugated".
I.e., the left (right) half of the chain
is conjugated while the other is unconjugated.
Explicitly, 
the multi-wave function overlap
(the triple inner product) for 
$\Psi_{\alpha}$, $\Psi_{\beta}$ and $\Psi_{\gamma}$
($\equiv \mathrm{MWO}(\Psi_{\alpha},\Psi_{\beta},\Psi_{\gamma})$)
is evaluated as
\begin{align}
\label{triple inner calc}
\mathrm{MWO}(\Psi_{\alpha},\Psi_{\beta},\Psi_{\gamma})
=
c_{\alpha\beta\gamma}.
\end{align}
Here, $c_{\alpha\beta\gamma}$
is 
the $U(1)$ phase associated with 
the
multiplication of the transition functions, 
$g_{\alpha\beta}g_{\beta\gamma} =
g_{\alpha\gamma} c_{\alpha\beta\gamma}$.
While we consider here three physically identical states, 
it is also possible to consider three physically distinct states
and their triple inner product \cite{shiozaki2023higher}.
The triple inner product extracts the topological invariant, 
the Dixmir-Douady class, 
for parameterized families of MPS.
See Ref.\ \cite{ohyama2023higher} for more details. 
Here, we apply the triple inner product for SPT phases in (1+1) dimensions. 

Let us consider $G$-symmetric SPT ground states. 
Here, $G$ represents unitary onsite symmetry.
Such SPT phases can be classified by the second group cohomology, $H^2(G, U(1))$
\cite{Chen_2013, Schuch_2011, Pollmann_2012}.
As before, we consider three invertible states that are physically equivalent.
They can be prepared from a given state $|\Psi_0\rangle$
by acting with global symmetry transformations $g_\alpha \in G$, 
$|\Psi_{\alpha} \rangle 
\equiv
|\Psi_{g_{\alpha}} \rangle
= U(g_\alpha) |\Psi_0\rangle$.
Here, $U(g)$ represents 
a linear representation of $g\in G$
acting on the physical Hilbert space.
The distinction between these states is rather subtle as $g_\alpha$'s are global symmetry transformations -- 
In terms of the MPS representations, these states are in different MPS gauges.
By taking $g_\alpha=g$, 
$g_\beta=1$, 
$g_\gamma=h^{-1}$
where $g,h\in G$,
\label{triple inner}
from Eq.\ \eqref{triple inner calc},
the triple inner product 
$\mathrm{MWO}(\Psi_g, \Psi_{0}, \Psi_{h^{-1}})$
is given by
\begin{align}
g^{\ }_{g,1}\,g^{\ }_{1,h^{-1}}\, 
g^{\ }_{h^{-1},g}
=
V_{g} V_{h} V_{(gh)^{-1}}
=
e^{i\phi_{g,h}}1.
\end{align}
Here, $V_g$ represents 
the group action on 
the virtual Hilbert space of MPS,
and 
$e^{i \phi_{g,h}}\in H^2(G,U(1))$
is the projective phase, 
$
V_{g}V_{h} = V_{gh}e^{i \phi_{g,h}}
$.

%

Thus, the triple inner product can be considered as a way to extract the topological invariant of (1+1)d SPT phases. See Ref.\ 
\cite{ohyama2024higher_a}
for further discussion between the SPT invariants (group cohomology phases) and the higher Berry phase 
(e.g., constant-rank, v.s. non-constant-rank
MPS).
We note that there are many other ways to extract the SPT invariant -- see, e.g., 
\cite{Haegeman_2012,kapustin2014symmetry,Shiozaki_2017,Kapustin_2017}.

The topological invariant can also be extracted in a slightly different manner.
Here, we consider three physically distinct states that 
are generated by partial symmetry transformations.
Let us consider a state 
$\Psi_{\alpha\beta}$ generated from some reference state (invertible ground state)
$|\Psi_0\rangle$
by acting partial symmetry operator $g_{\alpha}$ on the left half
and $g_{\beta}$ on the right half of the chain [Fig.\ \ref{fig:1+1d}(c)].
The partial symmetry operation generates a so-called mixed gauge MPS
[Fig.\ \ref{fig:1+1d}(a)].
We can then consider the triple inner product
as in Fig.\ \ref{fig:1+1d}(c).
These two inner products, written as
$\int \Psi_{\alpha}*\Psi_{\beta}* \Psi_{\gamma}$
and 
$\int \Psi_{\alpha\beta}*\Psi_{\beta\gamma}* \Psi_{\gamma\alpha}$
in Ref.\ \cite{ohyama2023higher},
are equal. 
Therefore, 
there are two different ways 
to extract the SPT invariants using 
the triple inner product --
the "global" and "partial" versions. 
Note that in the global version, non-trivial contributions come from spatial infinities, 
where the left- and right-eigenvectors of the transfer matrices are located.
Conversely, in the partial version, 
non-trivial contributions come from around the center of the trijunction.

\begin{figure*}[ht]
  \centering
  \includegraphics[scale=1.7]{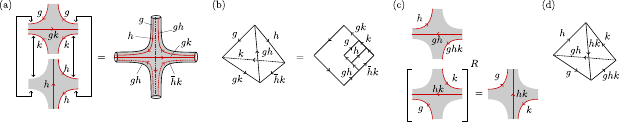}
  \caption{\label{fig:2+1d}
    The quadruple inner product in (2+1) dimensions.
    (a) The path integral for the quadruple inner product;
    the red lines represent symmetry twist defects
    (the arrows indicate $g$ or $g^{-1}$
    and do not represent the orientation of the surface.)
    (b) The defect line network for the sphere part of the path integral.
    (c) The relevant path integrals for the reduced density matrices, 
    with partial symmetry operations.
    The second path integral can be obtained by taking realignment.
    (d) The sphere part of the path integral that appears in the quadruple
    inner product.}
\end{figure*}

\subsection{Quadruple inner product in (2+1)d}
\label{Quadruple inner product in (2+1)d}

Let us now move on to the case of our main interest,
the quadruple inner product of (2+1)d invertible states.
The multi-wavefunction overlap in (2+1)d can be calculated by various means.
Ref.\ \cite{ohyama2024higher_a} uses
tensor-network representations of (2+1)d invertible ground states.
Here, we use the bulk-boundary correspondence
to reduce the calculation to some path integral in (1+1)d CFT. 
%


We first note that our wave function overlap can be expressed
in terms of the reduced density matrix $\rho_{AB}$ and its realignment $\rho_{AB}^R$,
More specifically, it can be expressed
a "mixed moment" of $\rho_{AB}$ and $\rho_{AB}^R$,
$\mathrm{Tr}\, [ (\rho_{AB}^R)^{\dag} \rho_{AB}]$.
Here, mixed moments are quantities like
$\mathrm{Tr}\, [(\rho_{AB}^R \rho^{\ }_{AB})^n]$
or
$\mathrm{Tr}\, [ ((\rho_{AB}^R)^{\dag} \rho_{AB})^n]$.
This type of mixed moment of the reduced density operators
appears when constructing the many-body $\mathbb{Z}_2$ topological invariant
of time-reversal symmetric states
\cite{Shapourian_2017,Shiozaki_2018, kobayashi202421dtopologicalphasesrt}.
As we will see, the corresponding path integral is topologically
four infinite cylinders connected with each other.
In addition, we consider introducing symmetry-twist defects in the path integral, as we will discuss momentarily.


Let us now consider four invertible many-body states
$\Psi_{\alpha},\Psi_{\beta}, \Psi_{\gamma}$ and $\Psi_{\delta}$ defined on an infinite 2d plane,
and the tripartition of the total space into
three regions $A$, $B$ and $C$ as in Fig.\ \ref{fig:2+1d}(a).
Following the (1+1)d case presented in Fig.\ \ref{fig1}(b),
we consider the pattern of contraction
depicted in Fig.\ \ref{fig1}(c).
Here, note that $\Psi_{\alpha}$ is unconjugated (ket),
$\Psi_{\beta}$ is conjugated (bra), 
and $\Psi_{\gamma}$ and $\Psi_{\delta}$ are "half conjugated".

As before, for SPT ground states,
different wavefunctions can be obtained by acting with symmetry,
$|\Psi_{\alpha}\rangle =|\Psi_{g_{\alpha}}\rangle = U(g_{\alpha}) |\Psi_0\rangle$.
The multi-wavefunction overlap above can be expressed by using 
``mixed'' density matrices and their realignment as follows.
We first consider the contraction of $\Psi_{\alpha}$ and $\Psi_{\beta}$,
which is represented, by taking partial trace over $C$,
as
$\mathrm{Tr}\, _C|\Psi_{\alpha}\rangle \langle \Psi_{\beta}|$,
which we call a mixed density operator obtained from $\Psi_{\alpha}$ and $\Psi_{\beta}$.
Similarly, we consider the mixed-density matrix of $\Psi_{\gamma}$ and $\Psi_{\delta}$,
$\mathrm{Tr}\, _C|\Psi_{\gamma}\rangle \langle \Psi_{\delta}|$,
and consider its realignment.
The above procedure leads to two path integrals with symmetry-twist defect lines
depicted in the left hand side of Fig.\ \ref{fig:2+1d}(a).
We then glue the two path integrals by taking the trace,
leading to the right hand side of Fig.\ \ref{fig:2+1d}(a).
This is the path integral representation of the quadruple inner product,
$\mathrm{MWO}(\Psi_{\alpha},\Psi_{\beta}, \Psi_{\gamma}, \Psi_{\delta})$.
Here we note that the symmetry-twist defects are introduced by 
the symmetry actions on the edge (boundary) Hilbert space of the SPT phase.
The symmetry actions are anomalous (they suffer from a 't Hooft anomaly)
and characterized by $H^3(G, U(1))$
\cite{CLW11,
Else_2014,
MGSC18,
Kawagoe_2021, 
kapustin2024anomalous, rubio2024classifying,Garre_Rubio_2023,rubio2024fractional}.
In the following, we set
$g_{\alpha}=g$,
$g_{\beta}= k$,
$g_{\gamma} = h$,
and $g_{\delta}=1$.

Note that, as we started with the (2+1)d invertible states on the infinite 2d
plane,
the corresponding edge CFT is defined on an infinite liner.
In the following, we introduce an infra red (IR) cutoff $L$ and consider
the edge CFT put on a finite interval of length $L$.
We will then take the limit $L/\beta \to \infty$.
In this limit, the four cylindrical regions
in Fig.\ \ref{fig:2+1d}(a) become thin.
We can insert the complete set of states $\sum_p \ket{p}\bra{p}$ to each long cylinder,
which is dominated by the vacuum $\approx \ket{0}\bra{0}$.
In other words, in this limit,
the end points of the interval should have no effect.
The Riemann surface then breaks up into a sphere and four cylinders.
(Alternatively, 
we could start with invertible states put on a spatial sphere.
In this case, 
the corresponding edge CFT is defined on a finite circle of circumference
($\equiv L$).
By taking the limit $L/\beta \to \infty$,
the Riemann surface breaks up into {\it two} spheres and four cylinders.
By disregarding one of the spheres and focusing on the other,
we can recover the quadruple inner product of
the invertible states on an infinite 2d plane.)

For the cylinders, we note that 
there are three parallel symmetry lines (defects) running along them -- see \cref{fig:2+1d}(a). 
Because we are only keeping $\bra{0}$ at each insertion, we get a constraint for a cylinder,
$
\prod_{i} g_i = 1 
$,
where the product is overall symmetry defects going along the cylinder.
Otherwise, the path integral vanishes.
(All cylinders in \cref{fig:2+1d}(a) satisfy this constraint.)

As our central interest is the phase part of the multi wavefunction overlap,
we consider the ratio
$
\mathrm{MWO}(\Psi_{\alpha},\Psi_{\beta},\Psi_{\gamma},\Psi_{\delta})/\mathrm{MWO}(\Psi_{0},\Psi_{0},\Psi_0,\Psi_0)
$.
In the ratio, the contributions from the cylinders cancel, and we
extract the contribution from the defect network on the sphere
depicted in Fig.\ \ref{fig:2+1d}(b). Explicitly, the ratio is given
in terms of the F-symbol ($\equiv F$) and the quantum dimensions
of the symmetry-twist defects ($\equiv d_g$)
\begin{align}
  \frac{
  \mathrm{MWO}(\Psi_{\alpha},\Psi_{\beta},\Psi_{\gamma},\Psi_{\delta})}
  {\mathrm{MWO}(\Psi_{0},\Psi_{0},\Psi_0,\Psi_0)}
  = [F_{gk}^{g, h, \bar{h}k}]_{gh, k} \sqrt{d_{gk} d_{g} d_{h} d_{\bar{h}k}}.
\end{align}
(For the details of symmetry-twist defects and the F-symbols, see,
for example, \cite{Fuchs_2002, Aasen_2016, Bultinck_2018, Chang_2019, Cheng_2020}.)
For symmetry twist defects, their quantum dimensions are $1$. 
To evaluate the diagram, we used four projections  $\ket{0}\bra{0}$ on twisted Hilbert spaces.
However, we can freely change the phase factor of each $\ket{0}$.
Thus, $\mathrm{MWO}(\Psi_{\alpha},\Psi_{\beta}, \Psi_{\gamma}, \Psi_{\delta})$ depends on the choice of four vacua.
This redefinition of the phase shifts $\mathrm{MWO}(\Psi_{\alpha},\Psi_{\beta}, \Psi_{\gamma}, \Psi_{\delta})$ by coboundary.
Therefore, the F-symbol (or more precisely
its class) defines the element of the group cohomology,
$[F_{gk}^{g, h, \bar{h}k}]_{{gh}, k}
=
\omega(g, h, k) \in H^3(G, U(1))$.

Several comments are in order:


-- While we focused on the phase part of the multi-wavefunction overlap,
the amplitude part can also be calculated by using techniques that we will
describe in the next section.
The amplitude part 
includes 
the area law term,
the corner contribution, and the OPE coefficient
-- see the next section for more details.
For the ratio we consider above, these contributions cancel.

--
While the calculation above
is carried out for non-chiral states, 
the same calculation goes through for chiral states, as far as they are invertible, 
e.g., integer Chern insulators.

-- Similar to the (1+1)d case, 
we can utilize partial symmetry operations.
We consider the two path integrals
in Fig.\ \ref{fig:2+1d}(a).
Here, $h, gh, ghk$  
are the labels for group elements.
The first path integral can be generated by 
first considering 
partial symmetry operation
$U_C ( (gh)^{-1})$ acting only on 
region $C$,
$U_{C}( (gh)^{-1} ) |\Psi\rangle$,
and then taking the partial trace over $C$ of the density operator (transition amplitude), 
$
\tilde{\rho}_{AB}
( (gh)^{-1}):=
\mathrm{Tr}_C\, [ 
U_{C}( (gh)^{-1} ) |\Psi\rangle
\langle \Psi|]$.
Subsequently, 
we act with partial symmetry operations 
on $A$ and $B$,
$
U_A(h)\, 
\tilde{\rho}_{AB}
( (gh)^{-1})\, 
U^{\dag}_{B}(ghk)
$.
The second path integral is obtained by realignment
Fig.\ \ref{fig:2+1d}(c).
%
This is the reason 
to consider the realignment of $\rho_{AB}$
-- it has the effect of putting a "vertical" defect. 
Gluing these two path integral surfaces
and taking the limit 
$L/\beta\to \infty$,
the path integral factories
into four cylinders and 
a sphere as before.
The sphere part is given by
the sphere with the defects running as
Fig.\ \ref{fig:2+1d}(d).
The tetrahedron gives the 
F-symbol or the group cohomology phase, 
$\omega(g,h,k)$.

-- Finally,
let us discuss the precise modeling of the trijunction.
Here, note that acting with partial symmetry operations {\it does not}
create a junction connecting the symmetry lines.
This may not be entirely obvious from the microscopic setup.
However, if we model the tripartition setup 
by the vertex state, this seems a natural conclusion. Whether this is a proper model or not 
can be scrutinized by comparing 
it with, e.g., tensor network calculations.
For example, 
in Ref.\ \cite{Siva_2022},
the tensor network representation 
of Levin-Wen type models
near the tripartition entangling boundaries 
is presented.
Looking at this network, 
nothing singular or special does seem to happen at the trijunction. 
Note, however, 
that we can "combine" these lines to create a junction.
In other words, we can somehow "move" the location of the junctions (vertices).

\begin{figure*}[ht]
  \centering
  \includegraphics[scale=1.7]{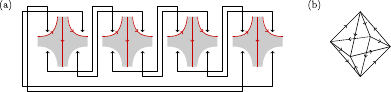}
  \caption{\label{fig:multi-entropy}
  (a)
  The path integral 
  representation of multipartition function $Z(A{\,:\,}B{\,:\,}C)$.
  (b)
  The spherical part of the multipartition function.
}
\end{figure*}

\begin{figure*}[ht]
  \centering
  \includegraphics[scale=1.7]{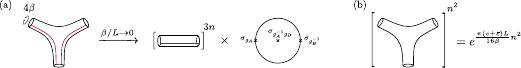}
  \caption{\label{fig:multi-entropy2}
  (a) The replica partition function for multi-entropy.
  The red lines represent $\mathbb{Z}_n$-like defects,
  along which different replicas are glued.
  (b) The path integral representing the normalization
  of multi-entropy.
 }
\end{figure*}

\section{multi-entropy} \label{sec:multi}

In this section, we consider multi-entropy for (2+1)d invertible states.

Multi-entropy
\cite{Gadde_2022, Penington:2022dhr,  Gadde:2023zzj, Harper:2024ker} 
is a quantity proposed to capture 
multi-partite entanglement.
For a $q$-partite ($q \geq 2$) pure quantum state $|\Psi\rangle$ 
defined on $q$ subsystems labeled by 
$K_1, K_2, \ldots, K_q$, 
the $n$-R\'enyi multi-entropy is defined using the replica trick as 
\begin{align}
    \label{multi-entropy definition}
    S^{(q)}_n = \frac{1}{1-n}\frac{1}{n^{q-2}} \ln
    \big(\langle \Psi|^{\otimes n^{q-1}} (\prod\nolimits_{j=1}^q \pi_{K_j}^{(j)}) |\Psi\rangle^{\otimes n^{q-1}}\big).
\end{align}
Here,
the permutation operators $\pi^{(j)}_{K_j}$ are in the group $S_{n^{q-1}}$ and permutes the $n^{q-1}$ copies of the subsystem $K_j$.
\footnote{
Here are some more details about the choice of permutation operators. First,
due to the symmetry of $|\Psi\rangle^{\otimes n^{q-1}}$,
we can always choose $\pi_{K_n}^{(n)} = e$. 
The remaining permutation operators 
are chosen such that  $\pi^{(j)}\equiv g^{(j)}$ 
generate 
$(\mathbb{Z}_n)^{\otimes q-1}$,
with $(g^{(j)})^n = e$ for $j = 1, \ldots, q-1$.
To ensure that each copy is treated equally in each permutation, it is required that all $\pi^{(j)} \in S_{n^{q-1}}$ have cycles of the same length, which means that each $\pi^{(j)}$ contains $n^{q-2}$ cycles of length $n$ for $1 \leq j \leq q-1$. We also note that such a definition of $\pi^{(j)}$'s ensures a symmetry among the $q$ subsystems, and such symmetry is expected to enhance the sensitivity of the multi-entropy to $q$-partite entanglement 
\cite{Gadde_2022, Harper:2024ker}.}
It is easy to see that $S^{(q)}_{n}$ vanishes for $q$-partite separable pure states.

We will mostly focus on
a special case of multi-entropy
with $q=3$ and $n=2$,
which is called $G(A{\,:\,}B{\,:\,}C)$ \cite{Penington:2022dhr}.
Specifically, for a pure tripartite quantum state $|\Psi\rangle$, 
we define
$G(A{\, :\,}B{\,:\,}C)$
and 
$Z(A{\, :\,}B{\,:\,}C)$ 
(``multi-partition function") 
as 
\begin{align}
\label{G_definition}
G(A{\, :\,}B{\,:\,}C) &= -\frac{1}{2}\ln Z(A{\,:\,}B{\, :\,}C),
\nonumber \\
Z(A{\, :\,}B{\,:\,}C) &= 
\langle \Psi|^{\otimes 4}(\pi_A^{(1)} \otimes \pi_B^{(2)} \otimes \pi_C^{(3)}) |\Psi\rangle^{\otimes 4}.
\end{align}
Here, the permutation operators $\pi^{(i)}$ are chosen to be
\begin{equation}
\pi^{(1)} = (1\:2)(3\:4), \; \pi^{(2)} = (1\:3)(2\:4), \; \pi^{(3)} = e,
\label{G_permutation}
\end{equation}
and $\pi^{(i)}_K$ permutes the 4 copies of the subsystem $K$ ($K = A, B, C$).

$G(A{\,:\,}B{\,:\,}C)$ has promising features to be a very sensitive measure of tripartite entanglement. 
For example, 
using the Cayley distance
as a distance measure in $S_4$, the three $\pi^{(i)}$'s are equidistant and incompatible.
Here, 
for two elements $\pi^{(i)}, \pi^{(j)} \in S_n$, the Cayley distance is defined as 
$d(\pi^{(i)}, \pi^{(j)}) = n - |C((\pi^{(i)})^{-1}\pi^{j})|$ where $|C(\pi)|$ is the number of disjoint cycles in $\pi$.
``Incompatible" means that there does not exist a $\pi$ such that $d(\pi, \pi^{(i)}) = d(\pi, \pi^{(j)})$ and $d(\pi, \pi^{(k)}) = d(\pi, \pi^{(l)})$ for $\{i, j\} \neq \{k, l\}$, $i, j, k, l = 1, 2, 3$. Moreover, it has been shown that for random tensor network states and holographic states, 
$G(A{\,:\,}B{\,:\,}C)$ is determined by minimal tripartitions of the system \cite{Penington:2022dhr}.

It is useful to note that
$G(A{\,:\,}B{\,:\,}C)$ is related to various other quantities measuring entanglement \cite{Penington:2022dhr, milekhin2022computablecrossnormtensor, PhysRevD.108.054508},
\begin{align}
    G(A{\,:\,}B{\,:\,}C) &= -\frac{1}{2} \mathcal{E}_{2}(AB) 
    \nonumber \\
    &= \frac{1}{2} S_{2, 2}^R(A{\,:\,}B) + S_2(AB).
    \label{G_CCNR_reflected entropy relation}
\end{align}
Here, ${\cal E}_n(AB)$ 
is the $n$-R\'enyi CCNR negativity,
\begin{align}
    \mathcal{E}_n(AB) 
    &= \ln \text{Tr}\,
    \big((\rho_{AB}^R(\rho_{AB}^R)^\dagger)^n\big),
    \label{CCNR_negativity_definition}
\end{align}
and $S_2(A)$ is the second R\'enyi entropy for subsystem $A$.
$S^R_{m,n}(A{\,:\,}B)$ is $(m,n)$-R\'enyi reflected entropy
-- see Appendix \ref{renyi reflected entropy} for the definition. This relation is useful for 
developing 
the correlation matrix method for calculating 
$G(A{\, :\,}B{\,:\,}C)$ 
for Gaussian states in 
Appendix \ref{appendix: covariance matrix approach for G(A:B:C)}.

We also note that
since R\'enyi entropies are nonnegative, Eq.\ \eqref{G_CCNR_reflected entropy relation} implies 
$G(A{\,:\,}B{\,:\,}C) - S_2(AB) \geq 0$. 
Since $S_2(AB) = S_2(C)$ for a tripartite pure state, 
$G(A{\,:\,}B{\,:\,}C) \geq S_2(C)$. 
Due to the invariance of 
$G(A{\,:\,}B{\,:\,}C)$ under the relabeling of the three subsystems~\cite{Penington:2022dhr}, 
we find that 
\begin{equation}
    G(A{\,:\,}B{\,:\,}C) \geq \max\{S_2(A), S_2(B), S_2(C)\}.
    \label{G(A:B:C) lower bound}
\end{equation}
Namely,
$G(A{\,:\,}B{\,:\,}C)$ is nonzero if there is any entanglement between the subsystems that is detectable by the $2$-R\'enyi entropy. 

To get some ideas as to 
how multi-entropy is calculated and 
what kind of quantum correlations
it captures,
let us consider 
simple examples, 
the GHZ and W states in a system of $N$ qubits:
\begin{align}
&
|\mathrm{GHZ}\rangle=
\frac{1}{\sqrt{2}}
\left(
|0\rangle^{\otimes N}+|1\rangle^{\otimes N}
\right),
\nonumber \\
&
|\mathrm{W}\rangle=\frac{1}{\sqrt{N}}
\left(|100 \ldots 0\rangle+|010 \ldots 0\rangle+\cdots+|00 \ldots 01\rangle
\right).
\end{align}
Now we partition the system into three parties $A$, $B$, and $C$, each containing $|A|$, $|B|$, and 
$|C|=N-|A|-|B|$ qubits and compute their 
multi-entropy. 
For the GHZ state, the realignment 
of the reduced density matrix in $A B$ is
\begin{equation}
\rho_{AB}^R = \frac{1}{2} \left(\ket{0}^{\otimes 2|A|} \bra{0}^{\otimes 2|B|} 
+ \ket{1}^{\otimes 2|A|} \bra{1}^{\otimes 2|B|} \right).
\end{equation}
Therefore the multi-partition function and 
multi-entropy are
\begin{equation}
Z_{\text{GHZ}}
= \frac{1}{8}, 
\quad
G_{\text{GHZ}} = \frac{3}{2} \ln 2.
\end{equation}
Surprisingly, the multi-entropy does not depend on $N$ or the tripartition.
For the W state,
\begin{align}
\rho^R_{AB} = \frac{1}{N} \Big[&(|100 \ldots 0\rangle+\cdots+|00 \ldots 01\rangle)
\nonumber \\
&\times
(\langle100 \ldots 0|+\cdots+\langle00 \ldots 01|)
\nonumber \\
&\quad + |C| |00 \ldots 0\rangle \langle 00 \ldots 0| \Big],
\end{align}
and therefore
\begin{equation}
Z_{\text{W}} = \frac{(|A|^2 + |B|^2 + |C|^2)^2}{N^4} ,
\quad
G_{\text{W}} = \ln \frac{N^2}{|A|^2 + |B|^2 + |C|^2}.
\end{equation}
In particular, for $N=3$, the canonical W state with $|A| = |B| = |C| = 1$ has
$G_{\mathrm{W}} = \ln 3$. We also note that the multi-entropy for the W state is symmetric under the exchange of the three subsystems $A$, 
$B$ and $C$.
In Appendix \ref{Appendix},
we provide more examples
of a few-body systems of anyons. 


Finally, it is useful to define the difference
\begin{align}
\label{def kappa}
\kappa \equiv 
G(A{\,:\,}B{\,:\,}C) - 
\frac{1}{2}
(S_2(A) + S_2(B) + S_2(C)).
\end{align}
This is similar to the Markov gap,
which is the difference between 
reflected entropy and mutual information \cite{Zou:2020bly, 2021JHEP...10..047H}.
For (2+1)d topological ground states,
the Markov gap is conjectured to be bounded from below 
by $((c+\bar{c})/3)\ln 2$ where $c+\bar{c}$ is the total central charge for ungappable degrees of freedom
\cite{Siva_2022, liu2022entanglement, liu2023multipartite}.
In the following, we will explore  
a similar universal bound for $\kappa$.


\subsection{CFT calculation}
\label{CFT calculation}

We now turn to the evaluation of 
multi-entropy for 
gapped ground states in (2+1) dimensions,
using the tripartition setup 
in Fig.\ \ref{tripartition setup}(a).
Starting from a trijunction state in \cref{tripartition setup}(b),
the path integral 
representation of
the multipartition function
$Z(A{\,:\,}B{\,:\,}C)
=
\operatorname{Tr}\,
[((\rho_{A B}^R)^{\dagger} 
\rho_{A B}^R)^2]$ 
is given by 
Fig.\ \ref{fig:multi-entropy}(a).
As with the multi-wavefunction overlaps, 
in the limit $L/\beta \to \infty$, we can insert 
the complete set of states
into each cylinder,
and then take the most dominant 
contribution from the vacuum.
This breaks the Riemann surface into a spherical part and six cylinders 
(two spherical 
parts when PBC is imposed).
The sphere(s) and cylinders
can be evaluated separately. 
By using the vacuum state $|0\rangle$ as the boundary condition for each cylinder,
in the $L/\beta \to \infty$ limit, each cylinder can be viewed as a torus with size $L \times 8\beta$. 
Thus, the contribution from each cylinder is 
$
\sim
\exp[\pi L (c+\bar{c})/96 \beta]$.
Combining 
the contributions from 
the six cylinders,
and taking into account
the normalization, 
the leading contributions to the multi-partition function 
and the multi entropy 
are
$
Z(A{\,:\,}B{\,:\,}C) \sim
\exp[
- 3 \pi L (c+\bar{c})/16\beta],
$
$G(A{\, :\,}B{\,:\,}C) \sim   
3 \pi L (c+\bar{c})/32\beta$.

In the above estimate, 
we only focus on the topological information 
when decomposing 
the path integral manifold into the cylinder and the sphere.
While capturing the leading contribution (the area law term in 
$G(A{\,:\,}B{\,:\,}C)$) correctly, 
it does not 
capture subleading ones,
which will be addressed below.

We now present a ``more accurate" calculation following 
the approach in \cite{liu2023multipartite}. 
First, we note that 
generic multi-entropy
can be represented as 
the replica partition function 
depicted in 
Fig.\ \ref{fig:multi-entropy2}(a).
Here, 
$\mathbb{Z}_n$-like symmetry defects
run along the time slices of 
the subregions $A$ and $B$.
Along these defect lines, different replicas are glued together. (The region $C$ is already traced out.)
Note that $n=2$ for our purpose.
In the limit $L/\beta\to \infty$, we decompose the 
path integral as shown
on the right-hand side of Fig.\ \ref{fig:multi-entropy2}(a).
Here, in the sphere part, twist operators
$\sigma_{g^{\ }_A}, \sigma_{g^{-1}_B}$,
and
$\sigma_{g_A^{-1}g^{\ }_B}$
are inserted. 
These twist operators, labeled by permutation operators $g_A$, $g^{-1}_B$, etc, permute replicas -- see 
\cite{Harper:2024ker} for the precise definition of the permutations.

In the earlier 
rough estimate,
during the decomposition into the cylinders and the sphere,
we implicitly unfolded the twist operator of the sphere and then thought of it as just an ordinary sphere.
While for $n=2$ the sphere with the twist operator is topologically equivalent to a standard sphere,
we must include the conformal factor when transforming it into a sphere through a conformal transformation.
A similar factor appears in the calculation of reflected entropy and the Markov gap
\cite{liu2023multipartite}.
The appearance of 
this factor can be understood by 
considering a simpler example: the
calculation of entanglement entropy for bipartition.
Recall that the replica partition function for 
entanglement entropy 
is created by inserting a $\mathbb{Z}_n$ symmetry defect into $\mathrm{tr}\, \rho_{AB}$.
(In our case, it is as if we are inserting a 
$\mathbb{Z}_n$ symmetry defect or its generalization into $\mathrm{tr}\, \rho_{ABC}$.)
As a result, the cylinder is twisted, and its period becomes 
$2\beta n$.
When decomposed using the identity operators,
the vacuum state in the 
$\mathbb{Z}_n$ “twisted” Hilbert space, i.e., the twist operator, dominates the sum.

The contribution from the 
twisted cylinders,
which consists of 
3 cylinders with periodicity 
$4n \beta$,
can be evaluated as
$
\sim 
\exp[ 
(\pi L (c+\bar{c})
/(12 \cdot 4n\beta) )
\times 3n
]
=
\exp[ \pi L ( c + \bar{c})/16 \beta]
$.
The sphere part with the twist operators can be evaluated as
\begin{align}
\left(\sin \frac{\pi}{3}\right)^{\frac{c+\bar{c}}{24}(n^2-1)\times 3}\times 
C_n
\end{align}
Here, $C_n \equiv C_{ \sigma_{g^{\ }_A} 
\sigma_{g_A^{-1}g^{\ }_B} 
\sigma_{g_B^{-1}} }$ 
is the operator product expansion (OPE) coefficient,
and
$(c+\bar{c})(n^2-1)/24$
is the dimension
of the twist operators
when $q=3$.

Finally, we also need to consider the normalization
given by Fig.\ \ref{fig:multi-entropy2}(b),
which originates from 
gluing the $n^2$ replicas 
trivially (without the twist operators).
Putting everything together, the multipartition function is given by
\begin{align}
 Z_n(A{\, :\,}B{\,:\,}C)
&= C_n\left(\sin \frac{\pi}{3}\right)^{\frac{c+\bar{c}}{8}\left(n^2-1\right)} 
e^{-\frac{\pi  L(c+\bar{c})}{16 \beta}\left(n^2-1\right)}. 
\end{align}
Correspondingly,
multi-entropy for $q=3$ is
\begin{align}
&
S^{(3)}_n(A{\,:\,}B{\, :\,}C)=  \frac{1}{1-n} \frac{1}{n} \ln Z _n
\nonumber \\
&\quad 
= \left(1+\frac{1}{n}\right) \frac{\pi L (c+\bar{c})}{16 \beta}
\nonumber \\
&
\qquad 
+
\left(1+\frac{1}{n}\right)
\frac{(c+\bar{c})}{8}
\ln \sin \frac{\pi}{3} 
+\frac{1}{1-n} \frac{1}{n} \ln C_n.
\end{align}
The term proportional 
to $\ln \sin \pi/3$
is the corner contribution
(or ``geometric contribution")
associated to 
the trijunction where
the regions $A,B,C$ meet.
Similar contributions appear 
also in 
reflected entropy 
\cite{liu2023multipartite}.

We now consider the difference
$\kappa_n \equiv 
S^{(3)}_n(A{\,:\,}B{\,:\,}C) - 
(1/2)
(S_n(A) + S_n(B) + S_n(C))
$
as in
Eq.\ \eqref{def kappa}.
Here, from \cite{liu2023multipartite},
the $n$-th R\'enyi entropy for subregion
$A$, which has a sharp corner with angle $2\pi/3$, 
is given by 
\begin{align}
S_n(A) &=
\left(1 + \frac{1}{n}\right)
\frac{\pi L(c+\bar{c})}{24 \beta}
\nonumber \\
&\quad
+
\left(1 + \frac{1}{n}\right)
\frac{(c+\bar{c})}{12}
\ln \sin \frac{\pi}{3}.
\end{align}
(Note however  
a slight difference between
the current setup 
and the one in 
\cite{liu2023multipartite}.
One of the main differences 
is that
in the setup of
\cite{liu2023multipartite},
one starts
with 
(2+1)d topological liquid
put on a spatial sphere,
and tripartition into 
three regions.
This setup has two trijunctions.
In the relevant CFT calculations,
one needs to 
consider CFT on a spatial circle,
and relevant path integrals
are defined on spacetime 
torii,
that can be reduced to cylinders
in the $\beta/L\to 0$ limit. 
Here, in this paper, 
we are considering open boundaries, and hence 
there is no need to consider a torus.) 
The area-law and the corner contributions
cancel, and $\kappa_n$
is finite,
\begin{align}
\kappa_n=
\frac{1}{1-n}\frac{1}{n}
\ln C_n.
\end{align}
While the explicit value of
$C_n$ is not known for generic $n$, when $n=2$,
\begin{align}
\label{main result}
\kappa = \frac{-1}{2}\ln 2^{-\frac{(c+\bar{c})}{4}}
=
\frac{(c+\bar{c})}{8}\ln 2.
\end{align}

In the above CFT calculation, 
we assume, among others,
that the entanglement spectrum is given 
by the spectrum of corresponding 
CFT --
see Sec.\
\ref{sec:Bulk-boundary correspondence and vertex states}.
If we move away from the ideal limit, 
following an analogy to 
the Markov gap, 
we expect that 
the equality in \eqref{main result}
is replaced by the universal bound,
$\kappa \ge ((c+\bar{c})/8)\ln 2$,
where $c+\bar{c}$ is the total central charge of ungappable 
degrees of freedom.
In other words,
the equality holds  
only when we remove all non-universal, 
short-range 
correlation
around 
the trijunction. 
Below, we will support this conjecture 
by the numerical calculation of 
$G(A{\,:\,}B{\,:\,}C)$
for the ground state of 
a lattice free-fermion model
realizing 
a (2+1)d Chern insulator phase
with chiral central charge $c=1$.

\begin{figure*}[ht]
\begin{tabular}{lll}
(a) & (b) & (c) \\
	\begin{minipage}[c][0.75\width]{0.18\textwidth}
	   \centering
    \includegraphics[scale=1.7]{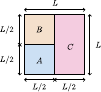}
	\end{minipage}
 &
	\begin{minipage}[c][0.75\width]{0.3\textwidth}
	   \centering
    \includegraphics[scale=0.3]{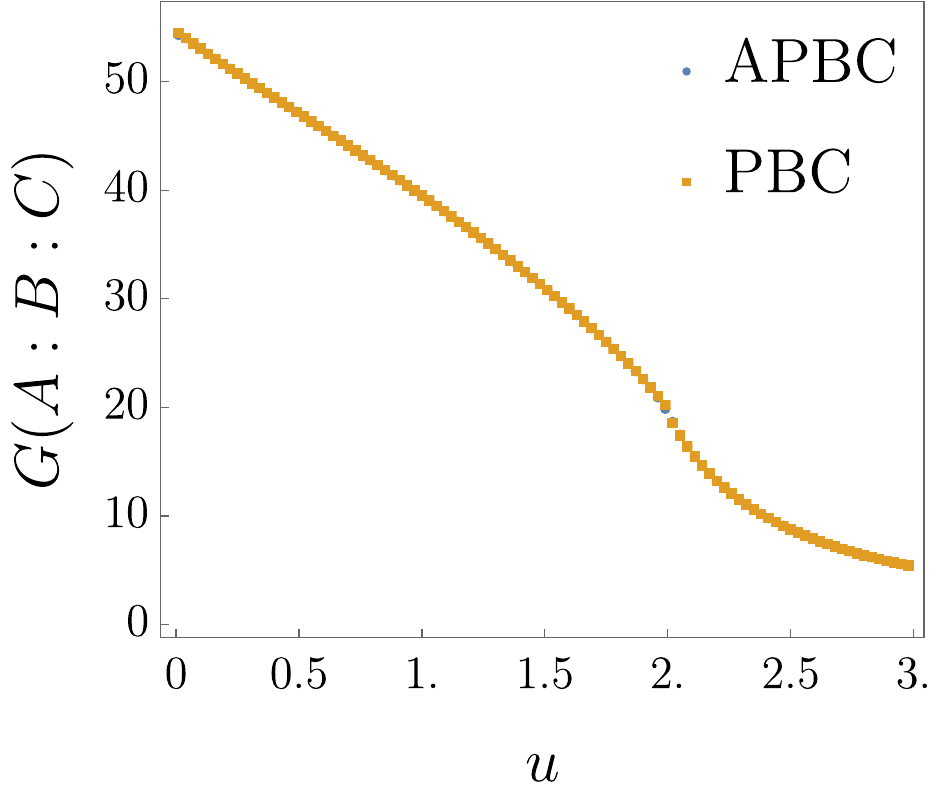}
	\end{minipage}
 &
	\begin{minipage}[c][0.75\width]{0.3\textwidth}
	   \centering
    \includegraphics[scale=0.3]{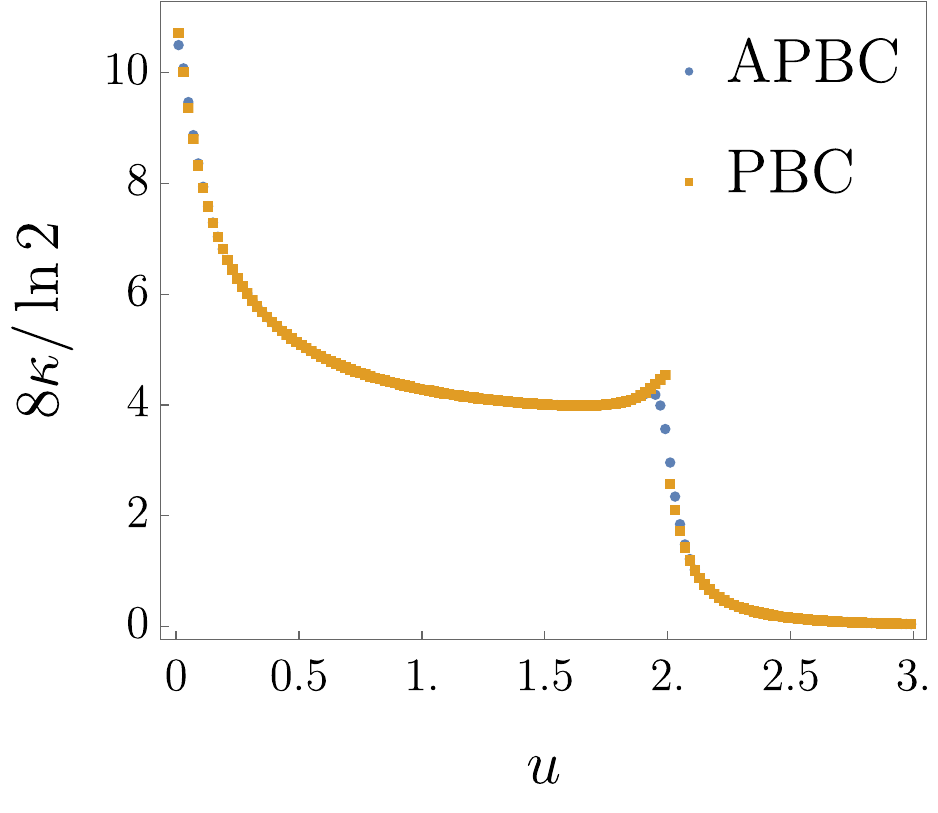}
	\end{minipage}
 \end{tabular}
    \caption{
\label{fig:G_kappa_vs_u_L_50}
(a) 
The tripartition of the $2$d lattice into three subsystems $A$, $B$ and $C$.
(b)
$G(A{\,:\,}B{\,:\,}C)$ 
and (c) $8\kappa/\ln 2$ 
for the Chern insulator model
on a $50 \times 50$ ($L = 50$) lattice, with anti-periodic boundary conditions (APBC) and with periodic boundary conditions (PBC). 
}
\end{figure*}

\begin{figure*}
\begin{tabular}{lll}
(a) & (b) & (c)\\
\includegraphics[scale=0.25]{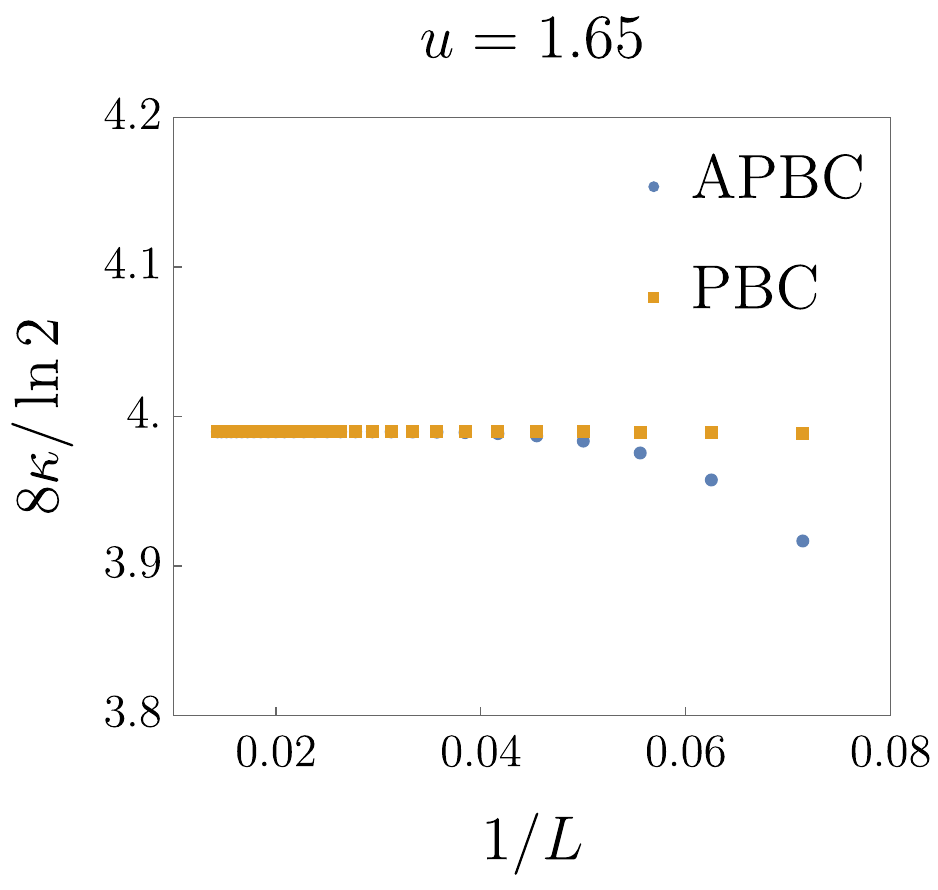}
\hspace{0.4cm}
&
\includegraphics[scale=0.25]{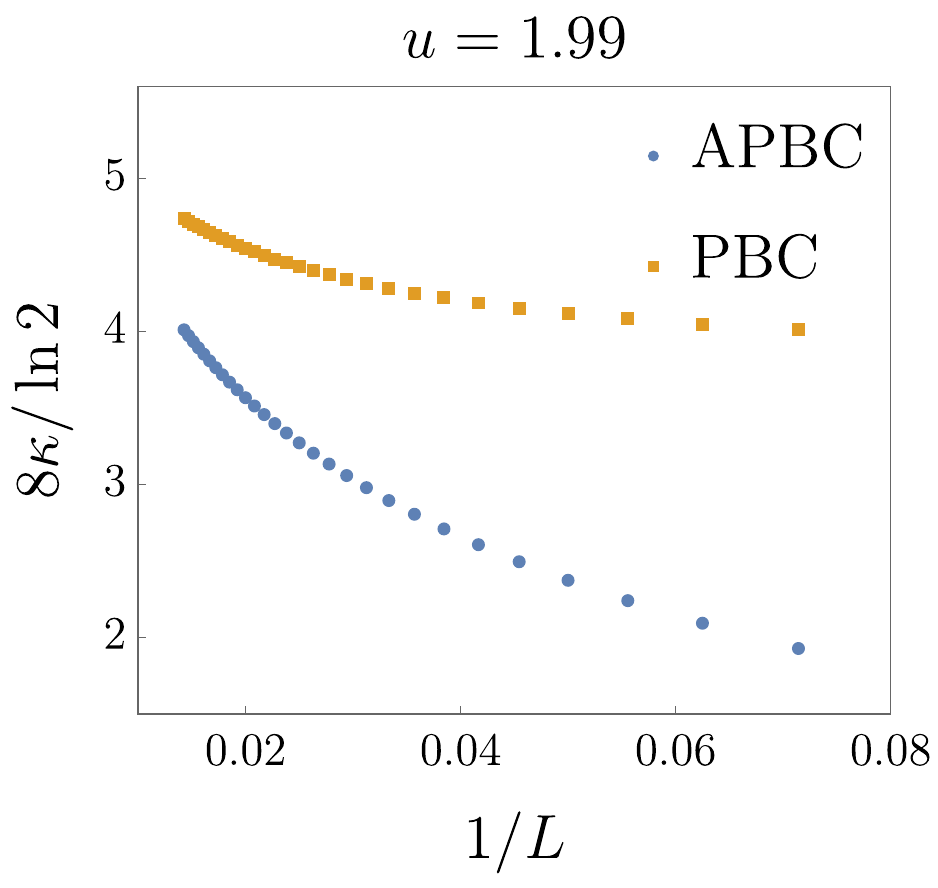}
\hspace{0.4cm}
&
\includegraphics[scale=0.25]{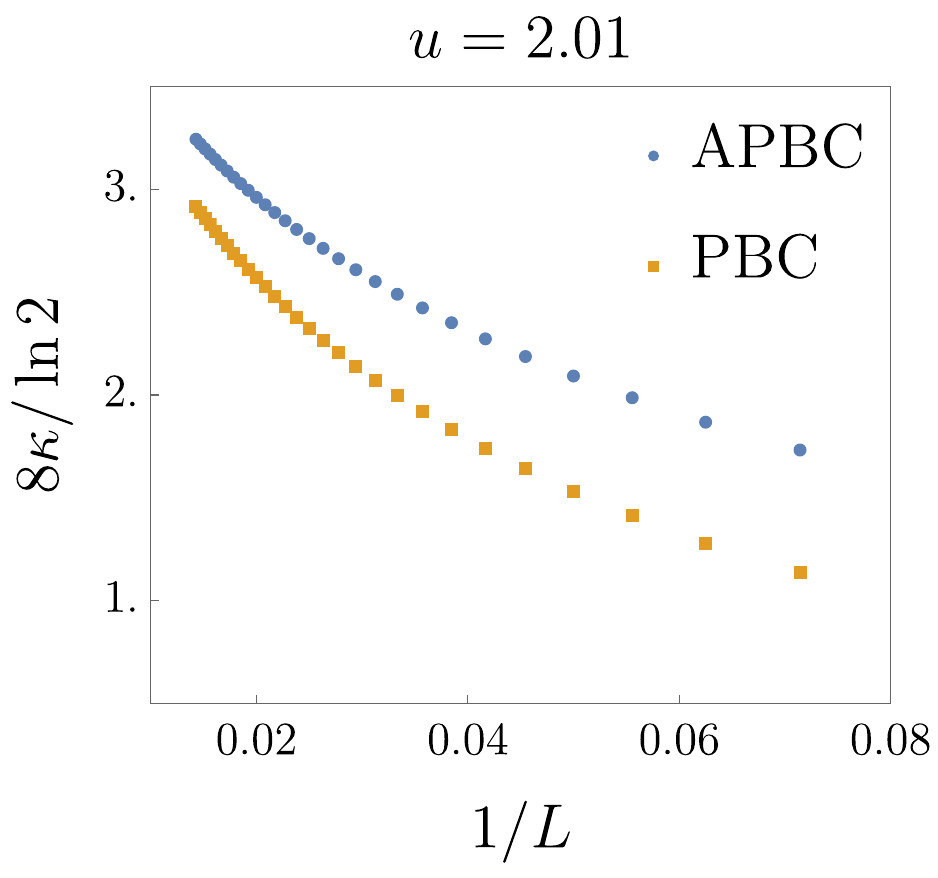}
\end{tabular}
    \caption{
\label{fig:kappa_min}
The scaling of $\kappa$ with inverse system size $1/L$ at 
$u = 1.65$ (a), $1.99$ (b), and $2.01$ (c) 
for 
the $(2+1)$d Chern insulator model. 
$u = 1.65$ is where $\kappa$ is 
the minimum in the topological phase with $\text{Ch} = 1$. 
}
\end{figure*}

\subsection{Numerics}
To study $G(A{\,:\,}B{\,:\,}C)$ and $\kappa$ for 
$(2+1)$d Chern insulator phases, 
we adopt the same lattice 
free fermion model
studied in 
Ref.\ \cite{Liu_2022}.
It is defined on a 2d square lattice
with the tight-binding Hamiltonian
\begin{equation}
\begin{split}
&
    H_{\text{Ch}} =  \frac{-i}{2}\sum_{\mathbf{r}}\sum_{i = x, y}[f_\mathbf{r}^\dagger \sigma_i f_{\mathbf{r}+\mathbf{a}_i} - f_{\mathbf{r}+\mathbf{a}_i}^\dagger \sigma_i f_\mathbf{r}] 
    \\ 
    &\quad + \frac{1}{2}\sum_{\mathbf{r}}
    \sum_{i = x, y}[f_\mathbf{r}^\dagger \sigma_z f_{\mathbf{r}+\mathbf{a}_i} 
    + f_{\mathbf{r}+\mathbf{a}_i}^\dagger \sigma_z f_\mathbf{r}] + u\sum_{\mathbf{r}}f_\mathbf{r}^\dagger \sigma_z f_\mathbf{r},
\end{split}
\end{equation}
where 
$f_{\mathbf{r}}$ represents 
a two-component fermion annihilation operator at cite $\mathbf{r}$, 
$\sigma_i$ $(i = x, y, z)$ are the Pauli matrices, and $\mathbf{a}_i$ is the unit vector along the 
$i=x,y$ direction. The model realizes
Chern insulator phases 
with Chern number $\text{Ch} = 1$ for $0 < u < 2$, $\text{Ch} = -1$ for $-2 < u < 0$, 
while it  
is in the trivial insulator phase with $\text{Ch} = 0$ otherwise. 
We place the system on a $L \times L$ square lattice with periodic
or antiperiodic boundary conditions. The lattice is tripartitioned such that the subsystems are rectangles with dimensions $L_{Ax} = L_{Bx} = L_{Cx} = L/2$ and $2L_{Ay} = 2L_{By} = L_{Cy} = L$. 
(The same tripartition setup 
as Ref.\ \cite{Liu_2022},
the setup
is shown in Fig.\ \ref{fig:G_kappa_vs_u_L_50}(a).)
The ground state $|\Psi\rangle_{ABC}$ is obtained by filling the lower half of the bands.


Similar to many entanglement quantities, 
$G(A{\, :\,}B{\, :\,}C)$ and $\kappa$ 
for Gaussian states (e.g., ground states of 
non-interacting Hamiltonians) can efficiently be 
calculated by the covariance matrix (correlation matrix) method. 
In Appendix \ref{appendix: covariance matrix approach for G(A:B:C)},
we outline our covariance matrix method.  
Plotted 
in Fig.\ \ref{fig:G_kappa_vs_u_L_50}(b,c)
are
numerically calculated
$G(A{\, :\,}B{\, :\,}C)$ and $\kappa$ 
for $L = 50$ 
and for $0 < u < 3$.
We see that $G(A{\,:\,}B{\,:\,}C)$ decreases monotonically as $u$ increases. 
On the other hand, 
$\kappa$ exhibits a peak around 
the phase transition point ($u=2$),
and approaches $0$ as we go deep into the topologically trivial phase. 
Similar behaviors are reported
for the Markov gap \cite{Liu_2022}. 


To check whether $\kappa$ takes the universal value in 
the Chern insulator phase, 
we look at the minimum of $\kappa$ in the $\text{Ch} = 1$ phase,
located around $u = 1.65$. 
We find that, for large enough $L$,
$\kappa$ is independent of system size and equals 
$\sim 3.990 \times ({c}/{8}) \ln 2$, as shown in Fig.\ \ref{fig:kappa_min}(a). This value is very close 
to 
the expected universal lower bound, 
$4 \times ({c}/{8}) \ln 2$,
discussed in Sec.\ \ref{CFT calculation}. 
We note that the extra factor of $4$ comes from the fact that the three subsystems meet in four junctions in the lattice model while they meet in only one junction in the CFT calculations \cite{Liu_2022}.


While we have focused on 
multi-entropy 
($G(A{\,:\,}B{\,:\,}C)$ and $\kappa$) for (2+1)d gapped 
topological ground states,
the numerical method we develop
allows us to discuss 
these quantities at (near) the phase transition. 
While $\kappa$ 
deep inside the gapped phases
is insensitive to 
the (sub)system sizes,
near the phase transitions, 
there is a strong finite size effect for 
$\kappa$
[Fig.\ \ref{fig:kappa_min}(b,c)]. 
Moreover, 
we see that the extrapolated 
values $\kappa$ 
at $u=1.99$ and $u=2.01$
for $1/L \to 0$ 
are still very different, signaling 
a diverging behavior
and
possible discontinuity in the continuum limit. 

\subsection{Defects}

As in the case of multi-wavefunction overlaps,
it is interesting to put defects (symmetry-twist defects) 
into the path integral
by (partial) symmetry transformations.
We note that 
partial symmetry transformations 
on the reduced density matrix 
can be utilized to define 
the so-called charged or grand canonical 
entanglement entropy
\cite{Belin_2013, Matsuura_2016}.
Its Fourier transform gives
the symmetry resolution of 
entanglement entropy
\cite{Goldstein_2018, Xavier_2018, Murciano_2020}. 
In the following, let us briefly discuss this generalization.
As before, let $G$ be the global symmetry 
of the system. Within the edge CFT, 
$G$ serves as the  
group of automorphism of 
the chiral algebra.
Applying symmetry transformation $a \in G$ to region $A$ and $c \in G$ to region $C$, and then apply the realignment operation (90-degree rotation), we 
obtain the path integral as in Fig.\ \ref{fig:multi-entropy}(a).
By gluing 
four copies of such 
reduced density matrices (with defect lines),
we define the multi-partition function as in Fig.\ \ref{fig:multi-entropy}(a).
The relevant path integral is represented 
in Fig.\ \ref{fig:multi-entropy}(a).
Taking the limit $L/\beta\to \infty$, 
the path integral surface again breaks into six cylinders and one spherical partition function with defect insertions. 
Each cylinder contains four parallel symmetry defect lines 
and $\ket{0}\bra{0}$ on each cylinder leg
enforces a constraint on 
the defect lines, 
such that they fuse to the identity.
On the spherical partition function, each $\ket{0}\bra{0}$ gives a vertex that connects four symmetry transformations. 
The symmetry transformations on the spherical partition function form an octahedron as shown in 
Fig.\ \ref{fig:multi-entropy}(b).
Since the diagram involves 4-vertex and different ways of breaking a 4-vertex into two 3-vertices are related by phase factors, the exact phase of is not well-defined. However, if we take periodic boundary condition instead of open boundary condition, the contribution from the conjugate diagram will cancel the phase
and the multi-partition function and 
multi-entropy will be real.

\section{Discussion}
\label{Discussion}

%
%
%

In this work, we developed an approach based on the bulk-boundary correspondence
to calculate the multi-wavefunction overlaps and multi-entropy for (2+1)d gapped states.
Specifically, the bulk-boundary correspondence enables us to reduce
these quantities to corresponding amplitudes (path integrals) 
in (1+1)d CFT possibly with defects.
The advantage of the approach is that it does not rely on microscopic representations 
of many-body ground states, such as tensor networks. 
While our primary focus was on invertible ground states, 
our method can potentially be extended to 
more complex cases such as non-invertible (topologically ordered) ground states. 

Regarding multi-entropy, we demonstrated that its behavior parallels 
that of reflected entropy and the Markov gap,
though our analysis in this paper
focused primarily on invertible ground states.
There remain many open questions regarding 
the aspects of many-body correlations
that multi-entropy captures.
For example, an immediate avenue to explore is 
which features of topological ground states are encoded by multi-entropy.
Similar to the Markov gap, 
one may speculate that multi-entropy is related 
to gappable edges of topologically ordered states.
We leave these questions to future studies.

More broadly, a deeper understanding of the connection 
between the non-zero value of $\kappa$
and the precise nature of multipartite entanglement,
in both topological ground states and beyond, is crucial.
It is clear that both $G(A{\,:\,}B{\,:\,}C)$ and reflected entropy vanish for product states of the form $|\Psi\rangle = |\psi_1\rangle_A \otimes |\psi_2\rangle_B \otimes |\psi_3\rangle_C$. 
Furthermore, for states with nontrivial entanglement between subsystems, 
both quantities are sensitive to how tripartite non-separable a pure state is. 
However, beyond these basic observations, 
it remains unclear what types of correlations $\kappa$ captures.

For the Markov gap, 
Ref.\ \cite{Zou:2020bly} proves a ``structure theorem",
which states that for a tripartite pure state, the Markov gap vanishes if and only if the state is 
a sum of triangle states (SOTS).
While $\kappa$ has been shown to vanish for single triangle states 
\cite{Penington:2022dhr},
it does not necessarily vanish for 
other SOTS.
More specifically, we can see that $\kappa=0$ for single triangle states by noting that $G(A{\,:\,}B{\,:\,}C) = ({1}/{2})(S_2(A)+S_2(B))$ 
for purely bipartite states $|\Psi\rangle_{ABC} = |\psi\rangle_{AB}|0\rangle_C$, and that $G(A{\,:\,}B{\,:\,}C)$ is additive under tensor products. 
Similar arguments can be applied to show that for single triangle states, $\kappa_n = 0$ for all integer $n \geq 2$, and thus the analytical continuation of $\kappa_n$ to $n = 1$ also vanishes for single triangle states
\cite{Penington:2022dhr}. 
On the other hand, for the GHZ state 
$\kappa_n = (\frac{n+1}{n} - \frac{3}{2})\ln 2$, which indicates that $\kappa_n$ with generic $n$ does not necessarily vanish for SOTS. Moreover, this example shows that similar to the R\'enyi generalization of the Markov gap, $\kappa_n$ with generic $n$ can take negative values. 
Finding structure 
theorems -- or lack thereof --
for $\kappa_n$ and related quantities 
remains an intriguing 
direction to explore.

%
%

%
%
%

\begin{acknowledgments}
S.R. thanks Jonathan Harper for useful discussion.
S.O. is supported by the European Union's Horizon 2020 research and innovation programme through grant no. 863476 (ERC-CoG SEQUAM) 
S.O. is also supported by JSPS KAKENHI Grant Number 23KJ1252 and 24K00522.
Y.K. is supported by the Brinson Prize Fellowship at Caltech and the U.S. Department of Energy, Office of Science, Office of High Energy Physics, under Award Number DE-SC0011632.
Y.K. is also supported by the INAMORI Frontier Program at Kyushu University and JSPS KAKENHI Grant Number 23K20046.
S.R. is supported 
by a Simons Investigator Grant from the Simons Foundation (Award No.~566116).
This work is supported by the Gordon and Betty Moore Foundation through Grant GBMF8685 toward the Princeton theory program. 
The authors thank the Yukawa Institute for Theoretical Physics at Kyoto University. Discussions during the YITP workshop YITP-T-24-03 on "Recent Developments and Challenges in Topological Phases" were useful to complete this work.
\end{acknowledgments}

\appendix 

\newcommand{\qTr}{\widetilde{\mathrm{Tr}}}
\newcommand{\trho}{\widetilde{\rho}}
\newcommand{\wt}{\widetilde}

\section{multi-entropy for anyon trimers}
\label{Appendix}

In this appendix, 
to gain further insights 
on the correlation measured by multi-entropy, 
we compute the multi-entropy for a normalized anyon trimer state 
in a theory with no fusion multiplicity \cite{Sohal_2023}
\begin{equation}
|\psi\rangle=\frac{1}{\left(d_a d_b d_c\right)^{1 / 4}}
\vcenter{\hbox{\begin{tikzpicture}[decoration={markings, mark=at position 0.5 with {\arrow{Stealth[length=1.mm,width=1.mm]}}}]
    \scriptsize
    \tikzset{mid arrow/.style={draw, thin, postaction={decorate}}}
    \draw[mid arrow] (0.6,1.2) -- (0,1.8) node[midway,left] {$a$};
    \draw[mid arrow] (0.6,1.2) -- (1.2,1.8) node[midway,right] {$b$};
    \draw[mid arrow] (1.2,0.6) -- (0.6,1.2);
    \draw[mid arrow] (1.2,0.6) -- (2.4,1.8) node[near end,right] {$c$};
\end{tikzpicture}}}
\end{equation}
Here,
$a,b,c,\cdots$ represent anyon labels and $d_{a,b,c,\cdots}$
are the corresponding quantum dimensions.
Here and henceforth,
we follow the notations in 
Ref.\ \cite{Sohal_2023}.
The reduced density matrix 
for anyons $a,b$ after tracing over anyon $c$ is
\begin{align}
    \trho_{AB} = \frac{1}{\sqrt{d_a d_b d_c}} 
\vcenter{\hbox{\begin{tikzpicture}[decoration={markings, mark=at position 0.5 with {\arrow{Stealth[length=1.mm,width=1.mm]}}}]
    \scriptsize
    \tikzset{mid arrow/.style={draw, thin, postaction={decorate}}}
    \draw[mid arrow] (0,0) -- (0.6,0.6) node[midway,left] {$a$};
    \draw[mid arrow] (1.2,0) -- (0.6,0.6) node[midway,right] {$b$};
    \draw[mid arrow] (0.6,0.6) -- (0.6,1.2) node[midway,left] {$c$};
    \draw[mid arrow] (0.6,1.2) -- (0,1.8) node[midway,left] {$a$};
    \draw[mid arrow] (0.6,1.2) -- (1.2,1.8) node[midway,right] {$b$};
\end{tikzpicture}}}
\end{align}
and the realignment reduced density matrix is 
\begin{align}
\trho_{AB}^R &= \frac{1}{\sqrt{d_a d_b d_c}} 
\begin{tikzpicture}[baseline={(current bounding box.center)}, decoration={markings, mark=at position 0.5 with {\arrow{Stealth[length=1mm,width=1mm]}}}]
    \scriptsize
    \draw[postaction={decorate}] (0,0) -- (0.6,0.6);
    \draw[postaction={decorate}] (1.2,0) -- (0.6,0.6) node[midway,right] {$b$};
    \draw[postaction={decorate}] (0.6,0.6) -- (0.6,1.2) node[midway,left] {$c$};
    \draw[postaction={decorate}] (0.6,1.2) -- (0,1.8) node[midway,left] {$a$};
    \draw[postaction={decorate}] (1.8,0) -- (1.8,1.2) node[midway,right] {$\bar b$};
    \draw[postaction={decorate}] (1.8,1.2) -- (1.2,1.8); 
    \draw[postaction={decorate}] (1.2,1.8) -- (0.6,1.2); 
    \draw (0,0) -- (-0.6,0.6) -- (-0.6,1.8) node[near end,left] {$\bar a$};
\end{tikzpicture}.
\end{align}
After an inverse F move and straightening out the legs,
$\tilde{\rho}_{AB}^R$ can be rewritten as
\begin{align}
& \trho_{AB}^R = \sum_f \frac{1}{\sqrt{d_a d_b d_c}} \left(F_{ab}^{ab}\right)^*_{fc}
\vcenter{\hbox{\begin{tikzpicture}[decoration={
    markings,
    mark=at position 0.5 with {\arrow{Stealth[length=1mm,width=1mm]}}}]
    \scriptsize
    \draw[postaction={decorate}] (0,0) -- (0,1.2);
    \draw[postaction={decorate}] (0,1.2) -- (0,1.8) node[left] {$a$};
    \draw[postaction={decorate}] (1.2, 0.6) -- (0, 1.2) node[midway,above] {$f$};
    \draw[postaction={decorate}] (1.2,0) -- (1.2,0.6);
    \draw (1.2, 0) node[right] {$b$};
    \draw[postaction={decorate}] (1.2,1.8) -- (1.2,0.6);
    \draw[postaction={decorate}] (1.8,1.2) -- (1.2,1.8);
    \draw[postaction={decorate}] (1.8,0) -- (1.8,1.2);
    \draw (1.8,0) node[right] {$\bar b$};
    \draw (0,0) -- (-0.6,0.6) -- (-0.6,1.8) node[left] {$\bar a$};
\end{tikzpicture}}} 
\nonumber \\
&\quad = \sum_f \frac{1}{\sqrt{d_a d_b d_c}} \left(F_{ab}^{ab}\right)^*_{fc} \left(A^{af}_a\right)^* B^{fb}_b
\vcenter{\hbox{\begin{tikzpicture}[decoration={markings, mark=at position 0.5 with {\arrow{Stealth[length=1.mm,width=1.mm]}}}]
    \scriptsize
    \tikzset{mid arrow/.style={draw, thin, postaction={decorate}}}
    \draw[mid arrow] (0,0) -- (0.6,0.6) node[midway,left] {$b$};
    \draw[mid arrow] (1.2,0) -- (0.6,0.6) node[midway,right] {$\bar b$};
    \draw[mid arrow] (0.6,0.6) -- (0.6,1.2) node[midway,left] {$f$};
    \draw[mid arrow] (0.6,1.2) -- (0,1.8) node[midway,left] {$\bar a$};
    \draw[mid arrow] (0.6,1.2) -- (1.2,1.8) node[midway,right] {$a$};
\end{tikzpicture}}}, 
\end{align}
where $A^{af}_a, B^{fb}_b$ 
are complex numbers of modulus one
and given by
\begin{equation}
A_c^{a b}=\sqrt{\frac{d_a d_b}{d_c}} \varkappa_a^*\left[F_b^{\bar{a} a b}\right]_{1,c}^* \ ,
\quad
B_c^{a b}=\sqrt{\frac{d_a d_b}{d_c}}\left[F_{\textcolor{red}{a}}^{a b \bar{b}}\right]_{c, 1}
\end{equation}
and $\varkappa_a$ is the Frobenius-Schur indicator of $a$.

The multi-partition function 
can be calculated as 
\begin{align}
& Z(A{\,:\,} B{\,:\,} C) 
=\operatorname{Tr}\left[\left(\left(\rho_{A B}^R\right)^{\dagger} \rho_{A B}^R\right)^2\right] 
\nonumber \\
&\quad = \sum_f \frac{1}{d_f} \qTr\left[\left(\left(\rho_{A B}^R\right)^{\dagger} \rho_{A B}^R\right)^2\right]_f 
\nonumber \\
&\quad = \sum_f \frac{\left|\left(F_{ab}^{ab}\right)_{fc} \right|^4}{d_f d_a^2 d_b^2 d_c^2} 
\vcenter{\hbox{\begin{tikzpicture}[decoration={
    markings,
    mark=at position 0.5 with {\arrow{Stealth[length=1mm,width=1mm]}}}]
    \scriptsize
    \draw[postaction={decorate}] (0,0) -- (0.6,0.6) node[midway,left] {$b$};
    \draw[postaction={decorate}] (1.2,0) -- (0.6,0.6) node[midway,right] {$\bar b$};
    \draw[postaction={decorate}] (0.6,0.6) -- (0.6,1.2) node[midway,left] {$f$};
    \draw[postaction={decorate}] (0.6,1.2) -- (-0.3,2.1) node[midway,left] {$\bar a$};
    \draw[postaction={decorate}] (-0.3,2.1) -- (-0.3,-5.7);
    \draw[postaction={decorate}] (0.6,1.2) -- (1.5,2.1) node[midway,right] {$a$};
    \draw[postaction={decorate}] (1.5,2.1) -- (1.5,-5.7);
    \draw[postaction={decorate}] (0,-1.8) -- (0.6,-1.2) node[midway,left] {$\bar a$};
    \draw[postaction={decorate}] (1.2,-1.8) -- (0.6,-1.2) node[midway,right] {$a$};
    \draw[postaction={decorate}] (0.6,-1.2) -- (0.6,-0.6) node[midway,left] {$f$};
    \draw[postaction={decorate}] (0.6,-0.6) -- (0,0) node[midway,left] {$b$};
    \draw[postaction={decorate}] (0.6,-0.6) -- (1.2,0) node[midway,right] {$\bar b$};
    \draw[postaction={decorate}] (0,-3.6) -- (0.6,-3.0) node[midway,left] {$b$};
    \draw[postaction={decorate}] (1.2,-3.6) -- (0.6,-3.0) node[midway,right] {$\bar b$};
    \draw[postaction={decorate}] (0.6,-3.0) -- (0.6,-2.4) node[midway,left] {$f$};
    \draw[postaction={decorate}] (0.6,-2.4) -- (0,-1.8) node[midway,left] {$\bar a$};
    \draw[postaction={decorate}] (0.6,-2.4) -- (1.2,-1.8) node[midway,right] {$a$};
    \draw[postaction={decorate}] (-0.3,-5.7) -- (0.6,-4.8) node[midway,left] {$\bar a$};
    \draw[postaction={decorate}] (1.5,-5.7) -- (0.6,-4.8) node[midway,right] {$a$};
    \draw[postaction={decorate}] (0.6,-4.8) -- (0.6,-4.2) node[midway,left] {$f$};
    \draw[postaction={decorate}] (0.6,-4.2) -- (0,-3.6) node[midway,left] {$b$};
    \draw[postaction={decorate}] (0.6,-4.2) -- (1.2,-3.6) node[midway,right] {$\bar b$};
\end{tikzpicture}}} 
\nonumber \\
&\quad = \sum_f \frac{\left|\left(F_{ab}^{ab}\right)_{fc} \right|^4}{d^2_f d_c^2} , 
\end{align}
where $\tilde{\mathrm{Tr}}$
represents the quantum trace.
Then the multi-entropy is given by
\begin{equation}
G(A{\,:\,}B{\,:\,}C) = - \frac{1}{2} \ln \sum_f \frac{\left|\left(F_{ab}^{ab}\right)_{fc} \right|^4}{d^2_f d_c^2}.
\end{equation}
As an example, let us consider the Ising anyon theory and take 
$a=\sigma, b=\sigma, c=\psi$. Then,
the multi-partition function 
is given by $Z=1/2$.
One can also check that 
taking
$a=\sigma, b=\psi, c=\sigma$ gives the same answer value, $Z=1/2$.
On the other hand, the multi-wave function overlap is calculated similarly as 
\begin{align}
\mathrm{MWO}\, (\psi,\psi,\psi,\psi) &= \operatorname{Tr}\, [(\rho_{A B}^R)^{\dagger} \rho_{A B}]
\nonumber \\
&= 
\frac{\delta_{a \bar{a}} \delta_{a b} }{d_c^2} \left(F_{aa}^{aa}\right)_{cc} A^{ac}_a \left(B^{cb}_b\right)^*.
\end{align}


\section{$G(A{\,:\,}B{\, }{\,:\,}C)$, R\'enyi CCNR negativity, and R\'enyi reflected entropy}
\label{renyi reflected entropy}

In this appendix, we provide a more detailed introduction to the quantities involved in Eq.\ \eqref{G_CCNR_reflected entropy relation} and their relations. 

First, we state the definition of (R\'enyi)  reflected entropy. 
For a mixed state $\rho_{AB}$ defined 
on a bipartite Hilbert space 
$\mathcal{H}_A\otimes \mathcal{H}_B$,
reflected entropy is defined as the von Neummann entropy 
of canonical purification.
More specifically,
for the (reduced) density matrix $\rho_{AB}$,
its canonical purification 
$|\rho_{AB}^{1/2}\rangle$
is defined in the doubled Hilbert space 
$\mathcal{H}_A\otimes \mathcal{H}_B\otimes 
\mathcal{H}_{A^*}
\otimes \mathcal{H}_{B^*}$
as
$|\rho_{AB}^{1/2}\rangle
= \sum_{a, a'}\sum_{b, b'} \langle ab|\rho_{AB}^{1/2}|a'b'\rangle |ab\rangle \otimes |a'b'\rangle^*$.
To define reflected entropy,
we take partial trace
over $\mathcal{H}_B\otimes \mathcal{H}_{B^*}$ 
and consider the reduced density matrix $\rho_{AA^*}$
as
\begin{equation}
    \rho_{AA^*} = \mathrm{Tr}_{\mathcal{H}_B \otimes \mathcal{H}_B^*}\,(|\rho_{AB}^{1/2} \rangle \langle \rho_{AB}^{1/2}|).
    \label{rho_AA}
\end{equation}
Finally, reflected entropy is defined as the von Neumann entropy of $\rho_{AA^*}$, 
\begin{equation}
    S^R(A{\, :\,}B) = -\text{Tr} \, (\rho_{AA^*} \ln \rho_{AA^*}).
    \label{reflected entropy definition}
\end{equation}


Reflected entropy can be generalized to a $(m, n)$-R\'enyi version ($m, n$ are positive integers). Consider the canonical purification $|\rho_{AB}^{m/2}\rangle$ of the the properly normalized density matrix 
\begin{equation}
    \rho_{AB}^{(m)} = \frac{\rho_{AB}^m}{\text{Tr}\, (\rho_{AB}^m)}.
    \label{rho_AB_m definition}
\end{equation}
The $(m, n)$-R\'enyi reflected entropy is defined as
\begin{equation}
    S_{m, n}^R(A{\,:\,}B) = \frac{1}{1-n} \ln \text{Tr}\,((\rho_{AA^*}^{(m)})^n)
    \label{Renyi reflected entropy definition}
\end{equation}
where $\rho_{AA^*}^{(m)}$ is similarly obtained from the density matrix of $|\rho_{AB}^{m/2}\rangle$ by tracing over $\mathcal{H}_B \otimes \mathcal{H}_B^*$,
\begin{equation}
    \rho_{AA^*}^{(m)} =  \mathrm{Tr}_{\mathcal{H}_B \otimes \mathcal{H}_B^*}\, (|\rho_{AB}^{m/2} \rangle \langle \rho_{AB}^{m/2}|).
    \label{rho_AA_m}
\end{equation}
Note that $\rho_{AB}^{(m)}$
can also be written in a similar form,  
$\rho_{AB}^{(m)}=  \mathrm{Tr}_{
\mathcal{H}_{A^*}\otimes 
\mathcal{H}_{B^*}}\, (|\rho_{AB}^{m/2} \rangle \langle \rho_{AB}^{m/2}|)$.

We now provide a more detailed discussion on the relations among 
$G(A{\,:\,}B{\,:\,}C)$, R\'enyi CCNR negativity, and R\'enyi reflected entropy. To relate R\'enyi CCNR negativity with R\'enyi reflected entropy,  we note the key relationship \cite{milekhin2022computablecrossnormtensor, PhysRevD.108.054508} that expresses $\rho_{AA^*}^{(2)}$ in terms of the realignment of $\rho_{AB}$
\begin{equation}
    \rho_{AA^*}^{(2)} = \frac{1}
    {\text{Tr}\, (\rho_{AB}^2)} \rho_{AB}^R (\rho_{AB}^R)^\dagger.
\end{equation}
Comparing with the definition of the $(2, n)$-R\'enyi reflected entropy \eqref{Renyi reflected entropy definition} and the definition of R\'enyi CCNR negativity  \eqref{CCNR_negativity_definition}, we see that $\mathcal{E}_n$ can be expressed as 
\begin{equation}
    \mathcal{E}_{n}(AB) = 
    (1-n)S_{2,n}^R(A{\, :\,}B) - nS_2(AB)
    \label{CCNR negativity as reflected entropy}
\end{equation}
where $S_2(AB)$ is the second R\'enyi entropy defined by $S_2(AB) = - 
\ln\text{Tr}\, (\rho_{AB}^2)$. In some sense, the $n$-R\'enyi CCNR negativity can be thought of as an unnormalized version of the $(2, n)$-R\'enyi reflected entropy
\cite{PhysRevD.108.054508}.

This equality can also be easily demonstrated from the replica trick definition of $\mathcal{E}_n$ and $S_{2,n}^R(A{\, :\,}B)$. Specifically, $\mathcal{E}_n$ is defined as \cite{Yin_2023a}
\begin{equation}
    \mathcal{E}_n(AB) = \ln 
    \text{Tr}\, 
    \big(\rho_{AB}^{\otimes 2n} ((\pi_n^{(1)})_A \otimes (\pi_n^{(2)})_B)\big)
    \label{replica trick definition for CCNR negativity}
\end{equation}
where the permutation operators acting on subsystems $A$ and $B$ are $\pi_n^{(1)} = (1 \: 2)(3 \: 4)\ldots (2n-1 \: 2n)$ and $\pi_n^{(2)} = (2 \: 3)(4 \: 5) \ldots (2n \: 1)$.
On the other hand, 
$S_{2,n}^R(A{\,:\,}B)$ 
can be written as
\cite{2021JHEP...03..178D}
\begin{align}
&
    S_{2, n}^R(A{\,:\,}B) 
    \nonumber \\
    &
    = \frac{1}{1-n}\ln
    \left[
    \frac{\langle \Psi|^{\otimes 2n} ((\pi_n^{(1)})_A \otimes (\pi_n^{(2)})_B)  |\Psi\rangle^{\otimes 2n}}{(\langle \Psi|^{\otimes 2} \tau_{AB} |\Psi\rangle^{\otimes 2})^n}
    \right],
    \label{replica trick definition for reflected entropy}
\end{align}
where $\pi_n^{(1)}$ and $\pi_n^{(2)}$ has the same definition as above and 
$|\Psi\rangle$ is defined on a system tripartitioned into subsystems $A$, $B$ and $C$. Thus, it is obvious that 
the numerator 
inside the logarithm
can be identified with the 2-R\'enyi CCNR negativity of the reduced density matrix $\rho_{AB}$. 
$\tau_{AB}$ acting on subsystem $AB$ is the cyclic permutation $(1 \: 2)$, and we identify the denominator with $-nS_2(AB)$. This shows how $\mathcal{E}_n$ can be expressed using the $(2, n)$-R\'enyi reflected entropy and the 2-R\'enyi entropy of $\rho_{AB}$.

Finally, we note that at $n=2$, $\pi_n^{(1)}$ and $\pi_n^{(2)}$ takes the exact same form as that in the definition of $\pi^{(1)}$ and $\pi^{(2)}$ in  $G(A{\,:\,}B{\,:\,}C)$ (see Eq.\ \eqref{G_permutation}). Combining this observation with Eq.\ \eqref{CCNR negativity as reflected entropy}, we arrive at Eq.\ \eqref{G_CCNR_reflected entropy relation}.

\section{The covariance matrix approach for $G(A{\,:\,}B{\,:\,}C)$}
\label{appendix: covariance matrix approach for G(A:B:C)}

In this appendix 
we develop the method of 
calculating $G(A{\,:\,}B{\,:\,}C)$ for non-interacting fermion systems 
in terms of the covariance matrix. 
Non-interacting fermion systems are described by Hamiltonians that contain only terms quadratic in the fermion creation and annihilation operators, and include free fermions and various topological phases of matter (e.g., Chern insulator, Kitaev chain, and more). The density matrix $\rho$ of a state in the Hilbert space of a non-interacting Hamiltonian is Gaussian \cite{Peschel:2002yqj}, i.e., it can be written as the exponential of a non-interacting fermion operator 
$\rho \propto e^{-H_E}$
where the entanglement Hamiltonian takes the form $H_E = \sum_{i, j} h_{i, j}^{(1)} f_i^\dagger f_j + h_{i, j}^{(2)} f_i^\dagger f_j^\dagger + H.c.$
We note that the reduced density matrix of a Gaussian density matrix is still Gaussian.

For a Gaussian (reduced) density matrix $\rho_{AB}$, all expectation values can be expressed in terms of two-point correlation functions $C_{ij} = \text{Tr}\, (\rho_{AB} f_i^\dagger f_j)$ and 
$F_{ij} = \text{Tr}\, (\rho_{AB} f_i^\dagger f_j^\dagger)$, and thus the state is completely determined by the correlation matrices $C = [C_{ij}]_{L \times L}$ (where $L$ is the system size) and $F = [F_{ij}]_{L \times L}$. 
In particular, the eigenvalues of $H_E$, and thus the entanglement entropy, can be calculated from the correlation matrices (or equivalently, the covariance matrices) \cite{Peschel:2002yqj, Vidal:2002rm, Peschel_2009}. 


\subsection{The covariance matrix in the Majorana fermion basis}
\label{covariance_matrix_introduction_subsection}

The calculation for $G(A{\,:\,}B{\,:\,}C)$ can be most conveniently performed in the Majorana fermion basis. Given complex fermion operators $f_j$ and $f_j^\dagger$ ($j = 1, \ldots, L$), the corresponding Majorana fermions are defined as $c_{2j-1} = f_j + f_j^\dagger$ and $c_{2j} = i(f_j - f_j^\dagger)$, which satisfy the anticommutation relation $\{c_{k}, c_{k'}\} = 2\delta_{kk'}$. The covariance matrix $\Gamma = [\Gamma_{kl}]_{2L \times 2L}$ is defined by 
\begin{equation}
     \Gamma_{kl} = \frac{1}{2} \text{Tr}\, (\rho_{AB}[c_k, c_l])
     \label{Covariance matrix definiton}
\end{equation}
and $\Gamma$ can be expressed in terms of the complex fermion correlation matrices as 
\begin{equation}
    \begin{split}
    \Gamma &=  (C - C^T) \otimes I_{2 \times 2} + (1 - C - C^T) \otimes \sigma_y \\ 
    &\quad + (F + F^\dagger) \otimes \sigma_z - i (F - F^\dagger) \otimes \sigma_x
    \end{split}
    \label{Covariance matrix in terms of correlation matrices}
\end{equation}
where $\sigma_x, \sigma_y$ and $\sigma_z$ are the $2 \times 2$ Pauli matrices acting on the space formed by neighboring odd- and even-indexed Majorana fermion operators $\{c_{2k-1}, c_{2k}\}$. 

By definition, $\Gamma$ is anti-symmetric. Since covariance matrices must also be Hermitian, $\Gamma$ is purely imaginary. Thus, $\Gamma$ can be block-diagonalized into 
\begin{equation}
    \Gamma = O^T 
    \left[\oplus_k \begin{pmatrix}
        0 & -i\gamma_k \\ i\gamma_k & 0
    \end{pmatrix}
    \right] O = O^T [\text{diag}(\gamma_k) \otimes \sigma_y] O
    \label{diagonalized covariance matrix}
\end{equation}
where $O$ is an orthogonal matrix. We define new Majorana fermion operators $c'_k = \sum_{j,k} O_{kj}c_j$. The only nonzero terms in the covariance matrix in the $c'_k$ basis are $\text{Tr}\,(\rho_{AB}[c'_{2k-1}, c'_{2k}]) = -i\gamma_k$ and $\text{Tr}\, (\rho_{AB}[c'_{2k}, c'_{2k-1}]) = i\gamma_k$ ($k = 1,\ldots,L$). From the correlation functions, we can obtain the form of the density matrix in the $c'_k$ basis
\begin{equation}
     \rho_{AB} = \prod_k
     \left[\frac{1}{2}(1+i\gamma_k c_{2k-1}' c_{2k}')
     \right]
     \label{density matrix in Majorana fermion basis}.
\end{equation}

From Eq.~\eqref{diagonalized covariance matrix}, we see that the eigenvalues of $\Gamma$ come in pairs of $\{\gamma_k, -\gamma_k\}$. 
The $n$-R\'enyi entropy for $\rho_{AB}$ can be expressed in terms of the eigenvalues of the covariance matrix as \cite{Vidal:2002rm, Peschel:2002yqj}
\begin{equation}
    S_n(AB) = \frac{1}{n-1} \sum_{k} \ln 
    \left[(\frac{1-\gamma_k}{2})^n + (\frac{1+\gamma_k}{2})^n\right].
    \label{n_Renyi_entropy_expression}
\end{equation}
Note that we only include one value in each pair of $\pm \gamma_k$ in the sum.

\subsection{Calculation of $G(A{\,:\,}B{\,:\,}C)$ from the covariance matrix method}
\label{G(A:B:C) calculation}

To calculate $G(A{\,:\,}B{\,:\,}C)$, we first calculate the $(2,2)$-R\'enyi reflected entropy $S_{2, 2}^R(A{\,:\,}B)$ which is related to $G(A{\,:\,}B{\,:\,}C)$ through Eq.\ \eqref{G_CCNR_reflected entropy relation}. The first step in this calculation is to find the covariance matrix for $\rho_{AB}^{(2)}$. 

Combining the definition of $\rho_{AB}^{(m)}$ in 
Eq.~\eqref{rho_AB_m definition} and the expression of $\rho_{AB}$ under the $c_k'$ basis in Eq.~\eqref{density matrix in Majorana fermion basis}, 
we find 
\begin{equation}
    \rho_{AB}^{(m)} = \prod_k \left[\frac{1}{2}(1 + i\frac{(1+\gamma_k)^m - (1-\gamma_k)^m}{(1+\gamma_k)^m + (1-\gamma_k)^m}c_{2k-1}'c_{2k}')
    \right].
    \label{rho_AB_m in Majorana fermion basis}
\end{equation}
Comparing with 
Eq.~\eqref{diagonalized covariance matrix}, we can see that the covariance matrix under the original $c_k$ basis takes the form
\begin{equation}
    \Gamma^{(m)} = O^T 
    \left[\text{diag}(\frac{(1+\gamma_k)^m - (1-\gamma_k)^m}{(1+\gamma_k)^m + (1-\gamma_k)^m}) \otimes \sigma_y
    \right] O.
    \label{Gamma_m}
\end{equation}
In particular, when $m = 2$, 
\begin{equation}
    \Gamma^{(2)} = O^T 
    \left[\text{diag}(\frac{2\gamma_k}{1+\gamma_k^2}) \otimes \sigma_y
    \right] O.
\end{equation}

For a state $\rho_{AB}^{(m)}$ with covariance matrix $\Gamma^{(m)}$, the covariance matrix for its purification $|\rho_{AB}^{m/2}\rangle$ can be expressed as \cite{PhysRevB.105.125125}
\begin{equation}
    \Gamma_{|\rho_{AB}^{m/2}\rangle} = \begin{pmatrix}
        \Gamma^{(m)} & \tilde{\Gamma}^{(m)} \\
        -\tilde{\Gamma}^{(m)} & -\Gamma^{(m)}
    \end{pmatrix}.
    \label{covariance matrix using TFD method}
\end{equation}
Here,
for a generic covariance matrix $\Gamma$ of the form $\Gamma = O'^T[\text{diag}(\eta_k)\otimes\sigma_y]O'$, $\tilde{\Gamma}$ is defined as $\tilde{\Gamma} = O'^T[\text{diag}(-i\sqrt{1-\eta_k^2}) \otimes I]O'$. Thus, given $\Gamma^{(m)}$,
\begin{equation}
    \tilde{\Gamma}^{(m)} = O^T \left[\text{diag}(\frac{-2i(1-\gamma_k^2)^{m/2}}{(1+\gamma_k)^m + (1-\gamma_m)^m}) \otimes I
    \right] O.
    \label{Gamma_tilde_m}
\end{equation}
For the specific case of $m = 2$, 
\begin{equation}
     \tilde{\Gamma}^{(2)} = O^T \left[\text{diag}(\frac{-i(1-\gamma_k^2)}{1+\gamma_k^2}) \otimes I
     \right] O.
     \label{Gamma_tilde_2}
\end{equation}

When $\rho_{AB}$ is Gaussian, $\rho_{AB}^{(m)}$ is also Gaussian since the product of Gaussian density matrices remains Gaussian. The canonical purification $|\rho^{1/2}\rangle$ is Gaussian for any Gaussian density matrix $\rho$ \cite{Bueno:2020vnx}, so $|\rho_{AB}^{m/2}\rangle$ is Gaussian, and the (R\'enyi) entropy of its reduced density matrices can be obtained from its covariance matrix. To calculate R\'enyi entropies corresponding to $\rho_{AA^*}^{(m)}$, we consider the covariance matrix $\Gamma_{|\rho_{AB}\rangle}$ restricted to subsystem $AA^*$, named $\Gamma_{AA^*}^{(2)}$, and denote its eigenvalues as $\{\pm \xi_l^{(2)}, l = 1,\ldots, 2L_A\}$.\footnote{Since $\rho_{AA^*}$ is defined in the doubled Hilbert space, the eigenvalues actually come in groups of 4 as $\{\xi_n^{(2)}, \xi_n^{(2)}, -\xi_n^{(2)}, -\xi_n^{(2)}\}$ with $1 \leq n \leq L_A$} 
We can express the $(2, 2)$-R\'enyi reflected entropy as
\begin{equation}
    S_{2, 2}^R(A:B) =  - \sum_{l} \ln 
    \left[(\frac{1-\xi_l^{(2)}}{2})^2 + (\frac{1+\xi_l^{(2)}}{2})^2
    \right]. \label{2_2_reflected_entropy_expression}
\end{equation}

Combining Eqs.\ 
\eqref{G_CCNR_reflected entropy relation},  \eqref{2_2_reflected_entropy_expression} and  \eqref{n_Renyi_entropy_expression} , we see that $G(A{\,:\,}B{\,:\,}C)$ takes 
the form
\begin{equation}
\begin{split}
    G(A{\,:\,}B{\,:\,}C) = & - \frac{1}{2} \sum_{l} \ln 
    \left[(\frac{1-\xi_l^{(2)}}{2})^2 + (\frac{1+\xi_l^{(2)}}{2})^2
    \right] \\ &- \sum_{k} \ln 
    \left[(\frac{1-\gamma_k}{2})^2 + (\frac{1+\gamma_k}{2})^2
    \right].
\end{split}
\end{equation}

As a side note, this procedure also allows us to calculate $S_{m, n}^R$ and $\mathcal{E}_n$ for generic $m, n$. 
Specifically, denoting the eigenvalues of $\Gamma_{AA^*}^{(m)}$ (the covariance matrix $\Gamma_{|\rho_{AB}^{m/2}\rangle}$ restricted to subsystem $AA^*$) as $\{\pm \xi_l^{(m)}, l = 1,\ldots, 2L_A\}$, the $(m, n)$-R\'enyi reflected entropy is expressed as 
\begin{equation}
    S_{m, n}^R(A{\,:\,}B) =  \frac{1}{1-n} \sum_{l} \ln 
    \left[(\frac{1-\xi_l^{(m)}}{2})^n + (\frac{1+\xi_l^{(m)}}{2})^n
    \right].
    \label{m_n_reflected_entropy_expression}
\end{equation}
According to Eq.\ \eqref{CCNR negativity as reflected entropy}, the $n$-R\'enyi CCNR negativity is then expressed as 
\begin{equation}
\begin{split}
     \mathcal{E}_{n}(AB) 
     &=  
      \sum_{l} \ln 
     \left[(\frac{1-\xi_l^{(2)}}{2})^n + (\frac{1+\xi_l^{(2)}}{2})^n\right] \\ 
     &\quad + 
     n\sum_{k} \ln 
     \left[(\frac{1-\gamma_k}{2})^2 + (\frac{1+\gamma_k}{2})^2
     \right].
     \label{n_CCNR_negativity_expression}
\end{split}
\end{equation}

\section{$\kappa$ and Markov gap in (1+1)d CFT}

In Sec.\ \ref{sec:multi}, we noted that $\kappa$ and the Markov gap share similar 
behaviors for (2+1)d topological ground states. 
In this Appendix, 
we briefly discuss these quantities 
for the ground state of (1+1)d CFT defined on a 1d ring of circumference $L$.
The ring is tripartitioned into three adjacent intervals $A, B, C$.

It has been shown \cite{Zou:2020bly, 2021JHEP...10..047H} 
that for $(1+1)$d CFTs with central charge $c$, the Markov gap $h(A{\, :\,}B)$ takes a universal value 
$h(A{\, :\,}B) = (c/3)\ln 2$, where in this section $c$ is the total central charge of the (1+1)d CFT. 
We note that we can define the $(m, n)$-R\'enyi generalization of the Markov gap, 
\begin{equation}
    h^{(m, n)}(A{\, :\,}B) = S^R_{m, n}(A{\, :\,}B) - I^{(n)}(A{\, :\,}B)
   \label{generalized Markov gap}
\end{equation}
where $I^{(n)}(A{\, :\,}B) = S_n(A) + S_n(B) - S_n(AB)$. We have numerically verified that it takes the value
\begin{equation}
    h^{(m, n)}(A{\, :\,}B) = \frac{1}{6}\frac{n+1}{n}\ln(2m)
    \label{generalized Markov gap_value}
\end{equation}
for $(1+1)$d free fermion CFTs.
This universal value 
can also be obtained from the OPE coefficient 
of the twist operators.
We also note that   
\cite{PhysRevD.108.054508} found by numerical simulations that $h^{(m, 1)}$ takes the value $({1}/{3})\ln(2m) - ({1}/{2})\ln(m)$ for 
the $(1+1)$d uncompactified free boson CFT. 

On the other hand, $\kappa$ is shown
to take the universal value 
$\kappa = ({c}/{4})\ln 2$ 
for the ground state of (1+1)d CFTs
\cite{Harper:2024ker}. 
We have numerically verified this using the covariance matrix method developed in Appendix \ref{appendix: covariance matrix approach for G(A:B:C)}. 

In Sec.\ \ref{CFT calculation}, we also examined the generalization of $\kappa$, named $\kappa_n$. We briefly mention its $(1+1)$d behavior here. 
It has been shown
\cite{Harper:2024ker} that in $(1+1)$d CFTs, 
\begin{equation}
    \kappa_n = \frac{1}{1-n}\frac{1}{n}\ln C_n
    \label{generalized_kappa_value}
\end{equation}
where $C_n \equiv C_{ \sigma_{g^{\ }_A} 
\sigma_{g_A^{-1}g^{\ }_B} 
\sigma_{g_B^{-1}} }$ is the OPE coefficient, and $\sigma_{g^{\ }_A}$, $\sigma_{g_B^{-1}}$ are the twist operators that appear in the expression of $G_n(A{\,:\,}B{\,:\,}C)$ 
as a three-point function. 
It is known that $C_2 = 4^{-4/c}$, but the explicit value of $C_n$ is not known for generic $n$. 

Finally, we note that $\kappa$ is at the point where the two series of values $h^{(m, n)}(A{\, :\,}B)$ and $\kappa_n$ coincide. More specifically, if we substitute in Eq.\ \eqref{G_CCNR_reflected entropy relation} for $G(A{\,:\,}B{\,:\,}C)$, and make the identification that $S_2(C) = S_2(AB)$ for tripartite pure states, we find that 
\begin{align}
    \kappa &= 
    \frac{1}{2}(S^R_{2, 2}(A{\, :\,}B) - I^{(2)}(A{\, :\,}B)) 
    \nonumber \\
    &= \frac{1}{2} h^{(2,2)}(A{\, :\,}B).
    \label{kappa_reflected_mutual_information}
\end{align}

\bibliography{ref}

\end{document}